\documentstyle[pra,aps,preprint,array]{revtex} 
\begin{document} 
 
\tighten \draft 
 
\title{Stochastic Dynamics of a Trapped Bose-Einstein Condensate} 
 
\author{R.A. Duine and H.T.C. Stoof} 
\address{Institute for Theoretical Physics, 
         Utrecht University, Leuvenlaan 4, \\ 
         3584 CE Utrecht, The Netherlands} 
 
\maketitle 
 
\begin{abstract} 
We present a variational solution of the Langevin field equation 
describing the nonequilibrium dynamics of a harmonically trapped 
Bose-Einstein condensate. If the thermal cloud remains in 
equilibrium at all times, we find that the equation of motions 
for the parameters in our variational {\it ansatz} are equivalent 
to the Langevin equations describing the motion of a massive 
Brownian particle in an external potential. Moreover, these 
equations are coupled to a stochastic rate equation for the 
number of atoms in the condensate. As applications of our 
approach, we have calculated the collisional damping rates and 
frequencies of the low-lying collective excitations of a 
condensate with repulsive interactions, and have obtained a 
description of the growth and subsequent collapse of a condensate 
with attractive interactions. We have found a good agreement with 
the available experimental results in both cases. 
\end{abstract} 
 
% Definitions 
\def\bx{{\bf x}} 
\def\bk{{\bf k}} 
\def\half{\frac{1}{2}} 
\def\args{(\bx,t)} 
 
\section{Introduction} 
The experimental realization of Bose-Einstein condensation in 
dilute atomic gases \cite{jila,curtis0,mit}, has led to a large 
increase in the amount of both experimental and theoretical 
research on these quantum systems. Various theoretical 
predictions regarding equilibrium and nonequilibrium properties 
of degenerate Bose gases, can now be directly compared with 
experimental data. Regarding the zero-temperature behaviour of a 
Bose-Einstein condensate, a great deal of the physics is well 
described by the Gross-Pitaevskii equation, i.e., a mean-field 
equation for the macroscopic wave function of the condensate. It 
has led to very good agreement with experimental results on, for 
example, the condensate collective mode frequencies and the 
density profile of the condensate at zero temperature \cite{gps}. 
To understand the nonzero temperature behaviour of Bose-condensed 
gases, several proposals have been made to generalize the 
Gross-Pitaevskii equation and to include the effects of the 
thermal cloud on the condensate. At the mean-field level this is 
achieved by introducing in the Gross-Pitaevskii equation real and 
imaginary terms, that describe the coherent and incoherent 
effects of collisions between condensate and thermal atoms, and 
that in particular cause evaporation or growth of the condensate 
\cite{dorfman,nick,henk1,walser,zgn}. However, at nonzero 
temperatures also fluctuations can play an important role. An 
example of this is the reversible formation of a condensate, as 
experimentally achieved by Stamper-Kurn {\it et al.} 
\cite{stamper}. Since the system is several times in the critical 
region, fluctuations of the order parameter around its mean-field 
value are of utmost importance to describe this experiment 
\cite{michiel1}. In addition, both quantum and thermal 
fluctuations are important to understand the stochastic nature of 
the collapse observed in $^7$Li 
\cite{collapse1,henk3,curtis,cass1,cass2,randy} and the phenomena 
of phase `diffusion' \cite{liyou}, in which case these 
fluctuations disturb the global phase of the condensate. Finally, 
from a fundamental point of view, a consistent description of a 
partially Bose-Einstein condensed gas requires that the 
fluctuation-dissipation theorem is obeyed, since this ensures 
relaxation of the system towards its correct physical 
equilibrium. Therefore, if dissipation is to be included in the 
generalized Gross-Pitaevskii equation, fluctuations must also be 
included. 
 
Gardiner and Zoller have included such fluctuations in the 
description of a Bose-condensed system by deriving with 
second-order perturbation theory a master equation for the 
one-body density matrix \cite{gardiner}, a procedure well-known 
from quantum optics. However, in this paper we will use the 
nonperturbative formulation developed previously by one of us 
\cite{henk1,henk2}. Using field-theoretical techniques Stoof 
derived a Fokker-Planck equation describing the full 
nonequilibrium probability distribution of the order parameter. 
An equivalent formulation of this theory can be given in terms of 
a dissipative nonlinear Schr\"odinger equation with noise. 
Although in principle we can turn to numerical methods for the 
solution of the Fokker-Planck equation, or its corresponding 
Langevin equation \cite{michiel1}, we find it more convenient 
here to proceed analytically, by means of a variational method. 
Variational approximations have previously provided a useful way 
to make analytical progress, and capture as many of the physics 
as possible. In particular, when applied to the zero-temperature 
Gross-Pitaevskii equation, a gaussian variational approximation 
has led to good results on the collective modes of the condensate 
\cite{gauss,michiel3,usama}, and on the description of the 
macroscopic tunneling of a condensate with attractive interactions 
\cite{henk3,cass1,legget,freire}. It is the aim of this paper to 
apply a similar variational method also to the dissipative 
nonlinear Schr\"odinger equation with noise appropriate for 
nonzero temperatures. We achieve this by assuming that the 
thermal cloud is in equilibrium at all times, and therefore acts 
as a `heat bath' on the condensate. The stochastic nonlinear 
Schr\"odinger equation then obeys an equilibrium version of the 
fluctuation-dissipation theorem, which ensures that the 
condensate relaxes to the physically correct equilibrium. With 
this assumption, we are then able to derive Langevin equations 
for the variational parameters in our gaussian {\it ansatz}, 
which turn out to be equivalent to the equations of motion for a 
Brownian particle in a potential. These equations are coupled to 
a stochastic rate equation for the number of atoms in the 
condensate. Using these equations of motion, we are then able to 
describe collisional damping of the condensate collective modes at 
nonzero temperatures, and the condensate growth and stochastic 
initiation of the collapse, as recently observed in $^7$Li 
\cite{cass2,randy}. 
 
The rest of this paper is organized as follows. To make this 
paper selfcontained, we review in Sec. II the techniques of path 
integrals, and their application to stochastic differential 
equations. We use the method of functional integration throughout 
this paper. In Sec. III we review the Fokker-Planck equation 
describing the nonequilibrium dynamics of a Bose-Einstein 
condensed gas, and discuss the equilibrium solution of this 
Fokker-Planck equation. The most important result of this section 
is the Langevin field equation for the order parameter, that obeys 
the fluctuation-dissipation theorem. This Langevin field equation 
takes the form of a dissipative nonlinear Schr\"odinger equation 
with noise. We also derive the stochastic equations of motion for 
the density and phase of the condensate, and a damped wave 
equation describing the propagation of sound waves in a 
homogeneous Bose gas, at nonzero temperatures. In Sec. IV we 
present the variational approximation to our nonlinear 
dissipative Sch\"odinger equation with noise, and also derive 
stochastic equations of motion for the variational parameters, 
which are the central result of this paper. To the best of our 
knowledge, a variational method for stochastic field equations, 
such as the stochastic non-linear Schr\"odinger equation under 
consideration here, has not been developed previously. In Sec. V 
we apply our equations to calculate the temperature dependence of 
the  damping and frequencies of the collective modes of a 
condensate, and to obtain a description of a growth-collapse 
curve of a condensate with attractive interactions. We end in 
Sec. VI with our conclusions.

\section{Path integrals and stochastic differential equations} 
In this section, we discuss the application of path integrals to 
stochastic differential equations. First, we consider Brownian 
motion in the overdamped limit, which means that the acceleration 
of the particle can be neglected with respect its damping rate. In 
the second part we consider underdamped Brownian motion, in which 
the nonzero mass of the particle plays a crucial role.

\subsection{Overdamped Brownian motion} 
Consider the one-dimensional Langevin equation \cite{nico,risken} 
\begin{equation} 
\label{1Dlangevin} 
  \dot q (t) = f (q(t)) + \eta (t). 
\end{equation} 
In this equation, $q(t)$ denotes the position of the particle, and $\eta (t)$ 
is a gaussian noise term, which has a time-correlation 
function given by 
\begin{equation} 
\label{timecor} 
  \langle \eta (t') \eta (t) \rangle = \sigma \delta (t'-t). 
\end{equation} 
Here, the brackets denote averaging over different realizations 
of the noise. The parameter $\sigma$ is positive and real, and is 
often referred to as the `strength' of the noise. For a given 
initial condition $q (t_0) = q_0$, the Langevin equation in 
Eq.~(\ref{1Dlangevin}) generates a probability distribution 
$P[q,t;q_0,t_0]$. By definition, $P[q,t;q_0,t_0] dq$ denotes the 
probability that a solution $q (t)$ of Eq.~(\ref{1Dlangevin}) 
starting at $q_0$ at time $t_0$, has a value between $q$ and 
$q+dq$ at time $t$. From this definition, we can immediately 
write down an expression for this probability distribution, i.e., 
\begin{equation} 
\label{defp} 
  P[q,t;q_0,t_0]=\langle \delta (q (t) - q ) \rangle. 
\end{equation} 
Here $q (t)$ denotes again a solution of the Langevin equation 
for a particular realization of the noise, starting at 
$q(t_0)=q_0$. The brackets denote averaging over different 
realizations of the noise, as in Eq.~(\ref{timecor}). 
 
We now want to derive a path-integral expression 
\cite{zinnjustin,kleinert} for the probability distribution 
$P[q,t;q_0,t_0]$. To achieve this, we first divide the time 
interval $t-t_0$ into $N$ pieces, each of length $\Delta = 
(t-t_0)/N$. Using the notation $q(t_n) \equiv q_n$, and 
$\eta(t_n)\equiv \eta_n$, we discretize the Langevin equation in 
Eq.~(\ref{1Dlangevin}) as follows 
\begin{equation} 
\label{langevindiscrete} 
  \frac{1}{\Delta} (q_{n+1}-q_n) = f(q_n) + \eta_n. 
\end{equation} 
The time-correlation of the noise is given by 
\begin{equation} 
\label{timecordiscrete} 
  \langle \eta_i \eta_j \rangle = \frac{\sigma}{\Delta} \delta_{ij}, 
\end{equation} 
which reduces to Eq.~(\ref{timecor}) in the limit $\Delta \to 0$. 
Making use of the fact that the noise has a gaussian 
distribution, we now write the probability distribution for a 
particular realization of the noise as 
\begin{equation} 
\label{noiseprobdiscr} 
  P(\{\eta_n\}) 
     = \left( \frac{2 \pi \sigma}{\Delta} \right)^{N/2} 
  \exp \left\{- \frac{\Delta}{2 \sigma} \sum_{n=0}^{N-1} \eta_n^2 
  \right\}. 
\end{equation} 
Using this probability distribution, we are able to calculate averages 
over the noise as gaussian integrals. The two-point function of the noise is, 
for example, given by 
\begin{equation} 
  \langle \eta_i \eta_j \rangle \equiv \int \prod_{n=0}^{N-1} d \eta_n 
   \eta_i \eta_j P(\{ \eta_n \}) , 
\end{equation} 
which reproduces the correct correlations, given in Eq.~(\ref{timecordiscrete}). 
We now first calculate the probability distribution $P[q_1,t_1;q_0,t_0]$, which is the 
probability distribution that a solution of the Langevin equation in Eq.~(\ref{1Dlangevin}) 
reaches the value $q_1$ at time $t_1 \equiv t_0 + \Delta$. 
Since we can easily solve the discrete Langevin equation in Eq.~(\ref{langevindiscrete}) 
explicitly for one time step, we find from the definition in 
Eq.~(\ref{defp}) that 
\begin{equation} 
  P[q_1,t_1;q_0,t_0] = \int 
         d \eta_0 \left( \frac{2 \pi \sigma}{\Delta} \right)^{1/2} 
     \exp \left\{-\frac{\Delta}{2 \sigma} \eta_0^2 \right\} \delta 
     \left(q_0 + \Delta (f(q_0)+\eta_0) - q_1\right). 
\end{equation} 
Secondly, we integrate out the noise $\eta_0$ to obtain the result 
\begin{equation} 
\label{probdeltat} 
  P[q_1,t_1;q_0,t_0] = \frac{1}{\Delta} \exp \left\{ -\frac{\Delta}{2 \sigma} 
     \left( 
       \frac{q_1-q_0}{\Delta} - f(q_0) 
     \right)^2 
   \right\}. 
\end{equation} 
We then use the this expression at each time step, and 
`tie' them together using 
\begin{equation} 
\label{probconserv} 
  P[q_{i+1},t_{i+1};q_{i-1},t_{i-1}] = \int d q_i 
    P[q_{i+1},t_{i+1};q_i,t_i] 
    P[q_{i},t_{i};q_{i-1},t_{i-1}], 
\end{equation} 
which follows from the fact that the total probability is conserved. 
The result for $P[q_N,t_N;q_0,t_0]$ then becomes, after a combination of 
Eqs.~(\ref{probdeltat}) and (\ref{probconserv}) at each intermediate time step, 
\begin{equation} 
\label{discrpathint} 
 P[q_N,t_N;q_0,t_0] = \Delta^{-(N-2)} 
  \int \left( \prod_{n=1}^{N-1} d q_n \right) \exp 
  \left\{ 
    -\frac{\Delta}{2 \sigma} 
    \sum_{i=0}^{N-1} \left( 
      \frac{q_{i+1}-q_i}{\Delta} - f(q_i) 
    \right)^2 
  \right\}. 
\end{equation} 
Note that this expression explicitly shows that the integration 
is only over intermediate coordinates, and that the boundary 
values $q_N$ and $q_0$ are fixed. We now take the limit $N \to 
\infty$ and $\Delta \to 0$, while keeping $t_N-t_0$ fixed. If we 
absorb the prefactor in Eq.~(\ref{discrpathint}) in the integral 
measure, we get, after putting $q_N=q$ and $t_N=t$, the result 
\begin{equation} 
\label{1Dprobpathint} 
  P[q,t;q_0,t_0] = \int_{q(t_0)=q_0}^{q(t)=q} d[q] e^{i S[q]/\hbar}. 
\end{equation} 
In the exponent of the integrant we extracted a factor $i/\hbar$, 
where $\hbar$ is Planck's constant, for reasons that become clear 
shortly. The integral measure of the functional integral in 
Eq.~(\ref{1Dprobpathint}) now denotes integration over all paths 
$q(t)$ with boundary conditions $q(t_0)=q_0$, and $q(t)=q$. Each 
of these paths gives a weighted contribution to the probability 
distribution, with a weight factor proportional to $e^{i 
S[q]/\hbar}$. The action $S[q]$ is given by 
\begin{equation} 
\label{actionq} 
  S[q]=i \hbar \int_{t_0}^t dt' \frac{1}{2 \sigma} 
  \left( 
    \dot q (t') - f(q(t')) 
  \right)^2. 
\end{equation} 
From the path-integral expression in Eq.~(\ref{1Dprobpathint}) we 
can now easily derive a partial differential equation for 
$P[q,t;q_0,t_0]$. This equation will turn out to be the 
Fokker-Planck equation \cite{nico,risken} corresponding to the 
Langevin equation of Eq.~(\ref{1Dlangevin}). To derive the 
Fokker-Planck equation, we make a connection with quantum 
mechanics. 
 
From ordinary quantum mechanics for a point particle with mass 
$m$ in a potential $V(q)$, we know that a matrix element of the 
evolution operator can be represented as a path integral 
\cite{zinnjustin,kleinert}, similar to Eq.~(\ref{1Dprobpathint}). 
Denoting that matrix element by $W(q,t;q_0,t_0)=\langle 
q|e^{-i(t-t_0) \hat H(\hat p, \hat q)/\hbar}|q_0\rangle$, where 
$\hat H(\hat p,\hat q)=\hat p^2/2 m + V(\hat q)$ is the Hamilton 
operator, this path integral is given by 
\begin{equation} 
\label{wpathint} 
  W(q,t;q_0,t_0)=\int_{q(t_0)=q_0}^{q(t)=q} d[q] 
  e^{i S^{{\rm cl}}[q]/\hbar}, 
\end{equation} 
with $S^{{\rm cl}}[q]$ the classical  action for the particle 
\begin{equation} 
\label{classaction} 
  S^{\rm cl}[q] = \int_{t_0}^t dt'\left( \frac{1}{2} m \dot q^2 (t')-V(q(t')) \right). 
\end{equation} 
On the other hand, the evolution operator also obeys the time-dependent Schr\"odinger equation 
\begin{equation} 
  i \hbar \frac{\partial}{\partial t} W (q,t;q_0,t_0) = H (p_q,q) W (q,t;q_0,t_0). 
\end{equation} 
Here, $H (p_q,q)$ is the hamiltonian corresponding to the 
classical action in Eq.~(\ref{classaction}), in the position 
representation. It is essential that $H (p_q,q)$ is normal 
ordered, i.e., that the momentum operators are positioned left to 
the position operators, for the path-integral expression in 
Eq.~(\ref{wpathint}) to be valid. We can now derive the 
Fokker-Planck equation by noting that Eq.~(\ref{1Dprobpathint}) 
can be interpreted as the path-integral expression for the time 
evolution operator with a `classical' action given by 
\begin{equation} 
  S[q] \equiv \int_{t_0}^t dt' L (t'), 
\end{equation} 
where $L(t)$ is by definition the lagrangian. The desired Fokker-Planck 
equation is, therefore, exactly the Schr\"odinger equation 
corresponding to the action $S[q]$. To derive the latter, we have to quantize this action, 
 which can be done in a straightforward manner. The momentum conjugate to the 
position variable $q$ is given by 
\begin{equation} 
  p_q \equiv \frac{\partial L}{\partial \dot q} = 
        \frac{i \hbar}{\sigma} (\dot q - f(q)). 
\end{equation} 
The hamiltonian thus becomes 
\begin{equation} 
\label{1Dfpham} 
  H(p_q,q) \equiv p_q \dot q - L (q,\dot q)= \frac{\sigma}{2 i \hbar} p_q^2 
  + p_q f(q). 
\end{equation} 
Note that we have positioned the momentum operators left to the 
position operators, in order to obtain the correct correspondence 
with Eq.~(\ref{1Dprobpathint}). Next, we impose canonical commutation 
relations $[q,p_q]= i \hbar$, so we write in the position 
representation $p_q=-i \hbar \partial/\partial q$. The 
`Schr\"odinger equation' for $P[q,t;q_0,t_0]$ is then given by 
\begin{equation} 
  i \hbar \frac{\partial P[q,t;q_0,t_0]}{\partial t} = H(p_q,q) 
  P[q,t;q_0,t_0], 
\end{equation} 
which can be explicitly rewritten as 
\begin{equation} 
\label{1Dfpeqn} 
  \frac{\partial P[q,t]}{\partial t} = 
  \frac{\partial}{\partial q} 
  \left[ 
    \frac{\sigma}{2} \frac{\partial}{\partial q} - f(q) 
  \right] P[q,t]. 
\end{equation} 
This is indeed the well-known Fokker-Planck equation corresponding to the Langevin 
equation in Eq.~(\ref{1Dlangevin}). 
It has a `streaming' term linear in the derivative with respect to $q$, which represents the 
reversible part of the Langevin equation. The 
`diffusion' term, quadratic in the derivatives, represents the irreversible 
stochastic behaviour. All Fokker-Planck equations that correspond to a 
Langevin equation with gaussian noise have this general structure. 
Note that in writing down the Fokker-Planck equation, we dropped the dependence 
of the probability distribution on the initial conditions. This is because the 
form of the Fokker-Planck equation is independent of these 
initial conditions. Although the hamiltonian in Eq.~(\ref{1Dfpham}) is not 
hermitian, 
the Fokker-Planck equation in Eq.~(\ref{1Dfpeqn}) 
conserves the normalization of the probability distribution, as can be seen 
from 
\begin{equation} 
 \frac{d}{dt} \int dq P[q,t] = \int dq \frac{\partial}{\partial q} 
  \left[ 
    \frac{\sigma}{2} \frac{\partial}{\partial q} - f(q) 
  \right] P[q,t] = \left.  \left[ \frac{\sigma}{2} \frac{\partial}{\partial 
  q} - f(q) 
  \right] P[q,t]  \right|^{+\infty}_{-\infty} = 0, 
\end{equation} 
where we made use of the fact that a normalized probability distribution 
has to vanish at the boundaries. 
The equilibrium solution of the Fokker-Planck equation is given by 
\begin{equation} 
  P[q,t \to \infty] \propto \exp \left\{ \frac{2}{\sigma} \int^q dq' f(q') 
  \right\}. 
\end{equation} 
Since we have used units such that  the mass of the Brownian 
particle and the friction coefficient are equal to one in the 
Langevin equation, we see that, in order for the probability 
distribution to relax to the correct Boltzmann distribution, we 
have to take the strength of the noise equal to $\sigma = 2 
k_{\rm B} T$, where $k_{\rm B}$ is Boltzmann's constant, and $T$ 
is the temperature. This is the fluctuation-dissipation theorem, 
which causes the probability distribution to relax to the correct 
physical equilibrium. 
 
At this point we would also like to make clear that in writing down the 
time sliced version of the Langevin equation, we 
have made the choice to interpret the noise term as a so-called Ito process, 
as opposed to a Stratonovich process. 
The difference between Ito and Stratonovich calculus emerges when one deals 
with multiplicative noise. For example, let us consider the Langevin 
equation 
\begin{equation} 
\label{1Dmultiplicative} 
  \dot q (t) = f(q(t)) + g(q(t)) \eta (t), 
\end{equation} 
where $\eta (t)$ is a gaussian noise term with correlations given by 
Eq.~(\ref{timecor}). 
If one interprets the noise in Eq.~(\ref{1Dmultiplicative}) as an Ito process, the 
discretization reads 
\begin{equation} 
  \frac{1}{\Delta} (q_{n+1}-q_n) = f(q_n) + g(q_n) \eta_n. 
\end{equation} 
The corresponding Fokker-Planck equation then 
becomes 
\begin{equation} 
 \frac{\partial P^{\rm I}[q,t]}{\partial t} = 
   \frac{\partial}{\partial q} 
   \left[ 
     \frac{\sigma}{2} \frac{\partial}{\partial q} g^2 (q) - f(q) 
   \right] P^{\rm I} [q,t]. 
\end{equation} 
However, the time sliced version of Eq.~(\ref{1Dmultiplicative}) is in the 
Stratonovich calculus given by \cite{nakazato} 
\begin{equation} 
  \frac{1}{\Delta} (q_{n+1}-q_n) = \frac{1}{2} [f(q_n) +f(q_{n+1})] 
     + \frac{1}{2} [g(q_n)+g(q_{n+1})] \eta_n. 
\end{equation} 
In this case, the Fokker-Planck equation reads 
\begin{equation} 
\label{stratfp} 
 \frac{\partial P^{\rm S}[q,t]}{\partial t} = 
   \frac{\partial}{\partial q} 
   \left[ 
     \frac{\sigma}{2} g^2 (q) \frac{\partial}{\partial q}  - f(q) 
   \right] P^{\rm S}[q,t]. 
\end{equation} 
The Stratonovich interpretation therefore leads to an additional noise-induced 
drift term in the equation for the average of $q(t)$, as can be seen from 
\begin{equation} 
\label{eomaverstrat} 
 \frac{d \langle q \rangle^{\rm S} (t)}{dt} = \langle f(q) \rangle^{\rm S} (t) + \sigma \langle 
   g (q)  g'(q) \rangle^{\rm S} (t), 
\end{equation} 
 where $g'(q) = \partial g/\partial q$. 
This result follows straightforward from the Fokker-Planck equation 
in Eq.~(\ref{stratfp}), with the use of partial integration. Note that the second term on 
the right-hand side of the last equation, which is the 
so-called spurious or noise-induced drift term, is absent in the case 
of an Ito process. In physics, a Stratonovich process 
arises naturally when the delta function in the time correlation of the noise is the 
result of a limiting procedure in which the correlation time becomes equal 
to zero. 
 
With these important remarks we conclude the discussion of the 
one-dimensional Langevin equation and its corresponding 
Fokker-Planck equation. As a further illustration of the 
path-integral methods, we now discuss the case of the Brownian 
motion of a massive particle in a potential. 
 
\subsection{Underdamped Brownian motion} 
Let us consider the Langevin equation for the Brownian motion in one 
dimension of a particle with mass $m$ in a potential $V(q)$, 
\begin{equation} 
\label{brownianeom} 
   \ddot q (t)+ \gamma \dot q (t) = - \frac{1}{m} \frac{\partial V}{\partial q} (q(t)) + \eta 
  (t). 
\end{equation} 
In this equation $\eta (t)$ is a fluctuating force per unit mass, with a gaussian 
probability distribution. The  parameter $\gamma>0$ is a friction constant. 
The time-correlation of the noise is given by 
\begin{equation} 
\label{flucforcecorr} 
  \langle \eta (t') \eta (t) \rangle = \frac{2 \gamma }{m \beta} \delta 
  (t'-t). 
\end{equation} 
Here $\beta = 1/k_{\rm B} T$ is the inverse temperature. Note 
that the strength of the fluctuations is related to the amount of 
dissipation $\gamma$ through Eq.~(\ref{flucforcecorr}). As 
mentioned previously, this is the fluctuation-dissipation 
theorem, which ensures that the probability distribution for the 
position and velocity of the Brownian particle relaxes to the 
Boltzmann distribution, as we will see in detail lateron. We now 
want to derive the Fokker-Planck equation associated with this 
Langevin equation of motion. To do so, we first write it as a set 
of two first-order differential equations 
\begin{eqnarray} 
\label{brownianfirstorder} 
  \dot q (t) &=& v(t); \nonumber \\ 
  \dot v(t) &=& - \gamma v(t) - \frac{1}{m} \frac{\partial V}{\partial q} (q(t)) 
                   + \eta (t). 
\end{eqnarray} 
We are interested in the probability distribution 
$P[q,v,t;q_0,v_0,t_0]$, which is defined as the probability density that a particle 
with velocity $v_0$ and position $q_0$ at an initial time $t_0$, has a 
velocity $v$ and a position $q$ at time $t$. This probability distribution is thus given 
by 
\begin{equation} 
\label{vandqprobdistr} 
  P[q,v,t;q_0,v_0,t_0] = 
  \langle 
    \delta (q(t) -q ) \delta (v(t)-v) 
  \rangle, 
\end{equation} 
where $(q(t),v(t))$ is a solution of 
Eq.~(\ref{brownianfirstorder}), with initial conditions 
$(q(t_0),v(t_0))=(q_0,v_0)$. Following the same time-slicing 
procedure as in the previous section, we can derive a 
path-integral expression for this probability distribution. This 
path integral is in first instance given by 
\begin{eqnarray} 
\label{pathintbrownqandv} 
  P[q,v,t;q_0,v_0,t_0] &=& 
   \int_{q(t_0)=q_0}^{q(t)=q} d[q] 
   \int_{v(t_0)=v_0}^{v(t)=v} d[v] \ 
   \delta [v (t') -\dot q (t')] \nonumber \\ 
   && \times \exp 
   \left\{ 
     -\frac{m \beta}{4 \gamma} 
     \int_{t_0}^t dt' 
     \left( 
       \dot v (t') + \gamma v (t') 
         + \frac{1}{m} \frac{\partial V}{\partial q} (q(t')) 
     \right)^2 
   \right\}. 
\end{eqnarray} 
We next  represent the delta functional by a Fourier path integral over 
 an auxiliary coordinate $p_q$. As the notation 
suggests, this turns out to be the momentum conjugate to $q$. The 
path-integral expression for $P[q,v,t;q_0,v_0,t_0]$ then becomes 
\begin{eqnarray} 
\label{pathinthreevars} 
  && P[q,v,t;q_0,v_0,t_0] = 
     \int_{q(t_0)=q_0}^{q(t)=q} d[q] 
     \int d[p_q] 
     \int_{v(t_0)=v_0}^{v(t)=v} d[v] \nonumber \\ 
     && \ \ \ \times \exp 
     \left\{ 
       \frac{i}{\hbar} \int_{t_0}^t dt' 
       \left( 
         p_q (t') (\dot q(t')-v(t')) 
         + \frac{i \hbar m \beta}{4 \gamma} 
           \left( 
             \dot v (t') + \gamma v (t') 
             +\frac{1}{m} \frac{\partial V}{\partial q} (q(t')) 
           \right)^2 
       \right) 
     \right\}. 
\end{eqnarray} 
In this expression, we again extracted a factor of $i/\hbar$ 
outside the time integral in the exponent to make the connection 
with quantum mechanics more obvious, and we also absorbed some 
normalization factors in the path-integral measure. We then 
introduce the momentum conjugate to $v$, denoted by $p_v$, by 
multiplying the integrant in Eq.~(\ref{pathinthreevars}) by a 
factor one, written as the gaussian functional integral over 
$p_v$ given by 
\begin{equation} 
  1 =  \int d[p_v] 
    \ \exp 
    \left\{ 
      \frac{i}{\hbar} 
      \int_{t_0}^t dt' 
      \frac{i \gamma}{\hbar m \beta} 
      \left( 
        p_v (t') - \frac{i \hbar m \beta}{2 \gamma} 
        \left( 
          \dot v(t') + \gamma v (t') 
          + \frac{1}{m} \frac{\partial V}{\partial q} (q(t')) 
        \right) 
      \right)^2 
    \right\}. 
\end{equation} 
This procedure is generally known as a Hubbard-Stratonovich transformation 
\cite{kleinert2}. 
After this procedure, the result for the probability distribution $P[q,v,t;q_0,v_0,t_0]$ reads 
\begin{eqnarray} 
\label{brownianresult} 
  P[q,v,t;q_0,v_0,t_0] &=& 
     \int_{q(t_0)=q_0}^{q(t)=q} d[q] 
     \int d[p_q] 
     \int_{v(t_0)=v_0}^{v(t)=v} d[v] 
     \int d[p_v] \nonumber \\ 
     && \times \exp 
     \left\{ 
       \frac{i}{\hbar} 
       \int_{t_0}^t dt' 
       \left( 
         p_q (t') q(t') + p_v (t') v(t') 
         - H (p_q,q;p_v,v) 
       \right) 
     \right\}, 
\end{eqnarray} 
with a hamiltonian given by 
\begin{equation} 
\label{fphambrown} 
  H(p_q,q;p_v,v) = p_q v - \frac{i \gamma}{\hbar m \beta} p_v^2 
    - p_v \left( \gamma v + \frac{1}{m} \frac{\partial V}{\partial q} \right). 
\end{equation} 
At this point it might be somewhat confusing that the momentum 
conjugate to $q$ is not simply proportional to $v$. We are, 
however, not quantizing a classical system, but instead trying to 
derive a path integral for the probability distribution generated 
by a classical stochastic equation of motion. The connection with 
quantum mechanics lies in the fact that we can identify this 
probability distribution with a quantum mechanical amplitude for 
some quantum system. This does not mean that the Brownian 
particle has wave-like properties. Nevertheless, 
Eq.~(\ref{brownianresult}) is precisely the canonical 
path-integral representation for a matrix element of the evolution 
operator. Therefore, the probability distribution 
$P[q,v,t;q_0,v_0,t_0]$ obeys the time-dependent Schr\"odinger 
equation with the hamiltonian $H (p_q,q;p_v,v)$ given by 
Eq.~(\ref{fphambrown}), in the position representation. It is 
again essential that we use normal ordering. We can thus quantize 
this hamiltonian, by putting $[q,p_q]=[v,p_v]=i \hbar$, with all 
other commutators equal to zero. So we have in the position 
representation $p_q=-i \hbar \partial/ \partial q$ and $p_v=-i 
\hbar \partial/ \partial v$. The Schr\"odinger equation now 
results in 
\begin{equation} 
\label{fpeqnbrownian} 
 \frac{\partial P[q,v,t]}{\partial t} 
   = 
  \left[ 
    -\frac{\partial}{\partial q} v 
    + \frac{\partial}{\partial v} 
      \left( 
       \gamma v + \frac{1}{m} \frac{\partial V}{\partial q} 
      \right) 
       + \frac{\gamma}{m \beta} \frac{\partial^2}{\partial v ^2} 
  \right] P[q,v,t], 
\end{equation} 
which is indeed the Fokker-Planck equation associated with the Brownian 
motion of a particle with mass $m$ in a potential $V(q)$, and known as the 
Kramers-Klein equation \cite{kramers}. It can be 
shown that the stationary solution of Eq.~(\ref{fpeqnbrownian}) is 
given by the Boltzmann distribution 
\begin{equation} 
  P[q,v,t \to \infty]  \propto 
  \exp 
  \left\{ 
    -\beta 
      \left( 
        \half m v^2 + V(q) 
      \right) 
  \right\}, 
\end{equation} 
which may be checked by insertion. It is important to realize 
that the fluctuation-dissipation theorem in 
Eq.~(\ref{flucforcecorr}) is again essential for the probability 
distribution to relax to the correct equilibrium distribution. It 
embodies the fact that dissipation and thermal fluctuations 
cooperate to achieve thermal equilibrium. This concludes our 
brief review of the connection between functional integration and 
stochastic differential equations. In the next section we use 
these techniques in the treatment of the nonequilibrium dynamics 
of a Bose-Einstein condensate.

\section{nonequilibrium dynamics} 
In this section, we present the Fokker-Planck 
equation describing the nonequilibrium dynamics of a Bose-Einstein 
condensed gas, and its corresponding Langevin field equation. The 
so-called hydrodynamic formulation will also be discussed. Since the Langevin 
field equation generalizes the Gross-Pitaevskii equation to nonzero 
temperature, we start our discussion by recalling this well-known equation. 
 
\subsection{Stochastic nonlinear Schr\"odinger equation} 
The dynamics of a trapped Bose-Einstein condensate is at sufficiently low 
temperatures very well described 
by the time-dependent Gross-Pitaevskii equation 
\begin{equation} 
\label{gpequation} 
   i \hbar \frac{\partial \Psi \args}{\partial t}= 
     \left\{ - \frac{\hbar^2 \nabla^2}{2m} 
             + V^{\rm{ext}}(\bx) 
         + T^{\rm 2B} |\Psi \args|^2 
     \right\}  \Psi \args, 
\end{equation} 
where $\hbar$ is Planck's constant, $m$ is the mass of a single 
atom, $V^{\rm{ext}} (\bx)$ is the external trapping potential and 
\mbox{$T^{\rm 2B}=4 \pi a \hbar^2/m$}  is the two-body transition 
matrix, with $a$ the s-wave scattering length. The 
Gross-Pitaevskii equation arises as the equation of motion for 
the superfluid order parameter, which is the expectation value of 
the Bose field operator $ \hat \psi \args$,  that annihilates an 
atom at position $\bx$ and at time $t$. The Gross-Pitaevskii 
equation is also referred to as the non-linear Schr\"odinger 
equation for the macroscopic condensate wave function $\Psi 
\args$, since the condensate density is given by 
\begin{equation} 
  n_c \args = |\Psi \args|^2. 
\end{equation} 
The time-dependent Gross-Pitaevskii equation has stationary 
solutions of the form $\Psi \args= \Psi (\bx) e^{i \mu t/\hbar}$, 
where the parameter 
$\mu$ is the chemical potential that fixes the number of atoms 
in the condensate and $\Psi (\bx)$ now obeys the time-independent 
Gross-Pitaevskii equation, 
\begin{equation} 
\label{gpeqntimeindep} 
   \left\{ - \frac{\hbar^2 \nabla^2}{2m} 
             + V^{\rm{ext}}(\bx) - \mu 
         + T^{\rm 2B} |\Psi (\bx)|^2 
     \right\}  \Psi (\bx) = 0. 
\end{equation} 
 
The time-dependent Gross-Pitaevskii equation is a semiclassical 
mean-field equation, describing the average dynamics of the 
condensate only. It contains no description of the relaxation of 
the condensate towards equilibrium, and neither does it contain 
condensate growth from the thermal cloud, or condensate 
evaporation, at nonzero temperatures. Moreover, it completely 
neglects fluctuations of the order parameter around its mean 
value in the description. Therefore, we would like to modify the 
Gross-Pitaevskii equation such that it contains fluctuations due 
to incoherent collisions between condensate and noncondensate 
atoms, as well as condensate growth and evaporation. In order to 
do so consistently, we have to consider the full probability 
distribution for the order parameter, which can be found by means 
of the many-body T-matrix approximation to a field-theoretic 
formulation of the Keldysh theory \cite{henk1,henk2}. It is given 
as a functional integral by 
\begin{equation} 
\label{probdistr} 
  P[\phi, \phi^*; t] = 
  \int^{\phi^* \args = \phi^* (\bx)}_{\phi \args = \phi (\bx)} 
    d[\phi^*] d[\phi] \nonumber \\ 
    \exp \left\{ \frac{i}{\hbar} S^{\rm{eff}} [\phi^*,\phi] 
    \right\}, 
\end{equation} 
with an effective action 
\begin{eqnarray} 
\label{seff} 
   S^{\rm{eff}} [\phi^*, \phi] && = \int_{t_0}^{t} dt' \int d \bx 
  \frac{2}{\hbar \Sigma^{\rm K} (\bx,t')} \nonumber \\ 
    &\times& \left| 
        \left( 
      i \hbar \frac{\partial}{\partial t'} +\frac{\hbar^2 \nabla^2}{2 m} 
      -V^{\rm{ext}} (\bx) + i R (\bx,t') + \mu (t')    +T^{\rm 2B} |\phi 
      (\bx,t')|^2 
    \right) \phi (\bx,t') 
     \right|^2. 
\end{eqnarray} 
In this effective action, the imaginary term $i R \args$ 
describes the exchange of atoms between the condensate and the 
thermal cloud. Since, at this point, we also want to be able to 
describe a thermal cloud that is not in thermal equilibrium, we 
have to allow for a time dependent chemical potential. Before we 
discuss the physical content of the expressions in 
Eqs.~(\ref{probdistr}) and (\ref{seff}) further, let us first 
derive the Fokker-Planck equation determining the time-dependence 
of $P[\phi, \phi^*; t]$. To do so, we note that the expressions 
in Eqs.~(\ref{probdistr}) and (\ref{seff}) are very similar to 
the path-integral expressions we encountered in the previous 
section for the probability distribution generated by a 
stochastic differential equation. The only difference between 
Eqs.~(\ref{1Dprobpathint}) and (\ref{probdistr}) is that the 
functional integration is now over all complex fields $\phi^* 
\args$ and $\phi \args$, instead of real functions $q (t)$. Also 
note that we did not specify the initial conditions at the time 
$t_0$. This is because we are only interested in the universal 
long-time dynamics of the gas, which are independent of the 
specific form of the initial conditions. Moreover, as we have 
seen in the previous section, the form of the Fokker-Planck 
equation is in fact independent of these initial conditions. From 
Eq.~(\ref{probdistr}) we can derive the Fokker-Planck equation by 
quantizing the effective action in Eq.~(\ref{seff}), just as in 
the previous section. It is ultimately given by 
\begin{eqnarray} 
\label{fpeqncondensate} 
  && i \hbar \frac{\partial}{\partial t} P [\phi^*,\phi;t] = \nonumber \\ 
  &&-\int d \bx \frac{\delta}{\delta \phi (\bx)} 
  \left( 
    - \frac{\hbar^2 \nabla^2}{2m} 
             + V^{\rm{ext}}(\bx) - \mu (t) 
         - i R\args 
         + T^{\rm 2B} |\phi (\bx)|^2 
  \right) \phi (\bx) P [\phi^*,\phi;t] \nonumber \\ 
  &&+\int d \bx \frac{\delta}{\delta \phi^* (\bx)} 
  \left( 
    - \frac{\hbar^2 \nabla^2}{2m} 
             + V^{\rm{ext}}(\bx) - \mu (t) 
         + i R\args 
         + T^{\rm 2B} |\phi (\bx)|^2 
  \right) \phi^* (\bx) P [\phi^*,\phi;t]  \nonumber \\ 
  &&-\half \int d \bx \frac{\delta^2}{\delta \phi (\bx) \delta \phi^* (\bx)} 
         \hbar \Sigma^{\rm K} \args P[\phi^*,\phi;t]. 
\end{eqnarray} 
This Fokker-Planck equation describes the time evolution of the 
probability distribution of the condensate wave function at 
nonzero temperatures, in the presence of a thermal cloud. 
 
The dissipation term $R \args$ describes the exchange of atoms 
between the thermal cloud and the condensate, due to elastic 
collisions. In the Hartree-Fock approximation, which is 
sufficiently accurate for the nonzero temperatures of interest 
here, it is given by \cite{henk2} 
\begin{eqnarray} 
\label{rterm} 
  R \args      & = & 2 \pi  \left( T^{\rm 2B}\right)^2 
                           \int \frac{d \bk_1}{(2 \pi)^3} 
                           \int \frac{d \bk_2}{(2 \pi)^3} 
               \int \frac{d \bk_3}{(2 \pi)^3} 
               (2 \pi)^3 \delta( \bk_1-\bk_2-\bk_3) 
                                  \nonumber \\ 
           &   & \times \delta \left( \epsilon +\epsilon_1 - 
                               \epsilon_2 -\epsilon_3 \right) 
            \left[ N_1 (1+N_2) (1+N_3) - (1+N_1) N_2 N_3  \right], 
\end{eqnarray} 
In this expression, $N_i \equiv N(\bx,\bk_i,t)$ is the Wigner 
distribution function of the thermal cloud, which can be 
determined by solving the corresponding quantum Boltzmann 
equation. We will not do this explicitly here, since lateron we 
assume that the noncondensed cloud is in thermal equilibrium. The 
energy of a thermal atom is given by 
\begin{equation} 
\label{energythermpart} 
  \epsilon_i = \frac{\hbar^2 \bk^2_i}{2 m} + V^{\rm{ext}} (\bx) 
              +  2 T^{\rm 2B} |\langle \phi (\bx) \rangle (t)|^2. 
\end{equation} 
Note that both in this expression, and in the Fokker-Planck 
equation in Eq.~(\ref{fpeqncondensate}) we neglected the effect of 
the mean field of the thermal atoms, because it plays a minor 
role in the dynamics of the condensate. Note also that, for the 
average value of the order parameter calculated with the 
probability distribution in Eq.~(\ref{probdistr}), we used the 
notation $\langle \phi (\bx) \rangle (t)$. The noisy order 
parameter field will be denoted by $\phi \args$ and for 
stochastic averages of this quantity we will use the notation 
$\langle \phi \args \rangle$. Since the Fokker-Planck equation 
and its corresponding Langevin equation are equivalent, we have 
of course that $\langle \phi (\bx) \rangle (t) =\langle \phi 
\args \rangle$. The Keldysh self-energy $\hbar \Sigma^{\rm K} 
\args$ in the Fokker-Planck equation 
 describes the thermal fluctuations due to incoherent collisions between 
 condensate and noncondensate  atoms. 
It is given explicitly by \cite{henk2} 
\begin{eqnarray} 
\label{sigmaktimedep} 
  \hbar \Sigma^{\rm K} \args & = & -4 \pi i \left( T^{\rm 2B}\right)^2 
                           \int \frac{d \bk_1}{(2 \pi)^3} 
                           \int \frac{d \bk_2}{(2 \pi)^3} 
               \int \frac{d \bk_3}{(2 \pi)^3} 
               (2 \pi)^3 \delta( \bk_1-\bk_2-\bk_3) 
                                  \nonumber\\ 
           &   & \times \delta \left( \epsilon +\epsilon_1 - 
                               \epsilon_2 -\epsilon_3 \right) 
                 \left[ N_1 (1+N_2) (1+N_3) + (1+N_1) N_2 N_3 \right]. 
\end{eqnarray} 
Note that both the dissipation $R \args$ and the Keldysh 
self-energy $\hbar \Sigma^{\rm K} \args$, depend on the energy $\epsilon$ to take 
a condensate atom out of the gas at position $\bx$ and time $t$, which has to be determined 
selfconsistently. This implies that $\epsilon$ 
is actually an operator in the configuration space of 
the order parameter, and given by \cite{henk2} 
\begin{equation} 
\label{hamgp} 
  \epsilon =-\frac{\hbar^2 \nabla^2}{2 m} +V^{\rm ext} (\bx) 
  +T^{\rm 2B} |\phi (\bx)|^2. 
\end{equation} 
The fact that $\epsilon$ should be viewed as an 
operator will turn out to be crucial for the probability 
distribution of the order parameter to relax to the correct 
equilibrium distribution function. 
 
Although our Fokker-Planck equation for the condensate, coupled 
to the appropriate quantum Boltzmann equation for the Wigner 
distribution function of the thermal cloud, describes in 
principle the full nonequilibrium dynamics of the Bose-condensed 
gas, its solution is very difficult even numerically. This is 
because of the fact that in the Fokker-Planck equation  the 
dissipation $R \args$ and the Keldysh self-energy also depend on 
the condensate wave function, through their dependence on 
$\epsilon$ given in Eq.~(\ref{hamgp}), and through the mean-field 
effect of the condensate on the thermal atoms. As a result, 
writing down the corresponding Langevin equation results in a 
stochastic equation with multiplicative noise, and with a 
prefactor of the noise that has a complicated dependence on $\phi 
\args$. We can, however, make progress by assuming that the 
thermal cloud is sufficiently close to equilibrium, which is, for 
example, justified for linear-response calculations around 
equilibrium, and also for condensate growth if the evaporative 
cooling is performed sufficiently slowly. From now on, we 
therefore assume that the thermal cloud can be described by a Bose 
distribution function 
\begin{equation} 
\label{bosedistr} 
  N (\epsilon_i) = \left[ e^{\beta (\epsilon_i-\mu)} - 
  1\right]^{-1}, 
\end{equation} 
with a chemical potential $\mu$ and an inverse temperature $\beta 
= 1/k_{\rm B} T$. The thermal cloud therefore now acts as a `heat 
bath' on the condensate. Making the $\epsilon$-dependence 
explicit for a moment, we can relate the dissipation $R 
(\bx;\epsilon)$, and the Keldysh self-energy $\hbar \Sigma^{\rm 
K} (\bx;\epsilon)$ by means of 
\begin{equation} 
\label{fdtheoremnoapprox} 
   i  R (\bx;\epsilon) =  -\half \hbar \Sigma^{\rm K} (\bx;\epsilon) [1+2 N(\epsilon)]^{-1}, 
\end{equation} 
which follows simply from the form of the Bose distribution function, 
together with the energy conserving delta 
function in Eqs.~(\ref{rterm}) and (\ref{sigmaktimedep}). 
This relation between the dissipation $R (\bx; \epsilon)$ and the 
 Keldysh self-energy $\hbar \Sigma^{\rm K} (\bx; \epsilon)$ determining the strength 
 of the fluctuations,  is in fact the 
fluctuation-dissipation theorem. Just as in the case of the 
Brownian motion of a particle discussed in the previous section, 
it causes the system to relax to the correct equilibrium 
distribution, as we will see below. Since we are dealing with 
Bose condensation, the occupation numbers $N(\epsilon)$ are 
generally very large, and we have in a good approximation 
\begin{equation} 
\label{approxtofdt} 
  [1+2N(\epsilon)]^{-1} \simeq \half [\beta (\epsilon - 
  \mu)]. 
\end{equation} 
If we combine this result with Eq.~(\ref{fdtheoremnoapprox}), and substitute the operator 
in Eq.~(\ref{hamgp}),  we 
arrive at the approximation 
\begin{equation} 
\label{fdtheorem} 
  i R \args \simeq -\frac{\beta}{4} \hbar \Sigma^{\rm K} 
  \args \left[ 
    -\frac{\hbar^2 \nabla^2}{2 m} +V^{\rm ext} (\bx) - \mu 
  +T^{\rm 2B} |\phi (\bx)|^2 \right], 
\end{equation} 
where $\hbar \Sigma^{\rm K} \args \equiv \hbar \Sigma^{\rm K} 
(\bx;\langle \mu_{\rm c} \args \rangle)$, and the local chemical 
potential of the condensate $\langle \mu_{\rm c} \args \rangle$ is 
given by 
\begin{equation} 
\label{chempotcondfunctional} 
  \langle \mu_{\rm c} \args \rangle= \frac{\delta}{\delta |\langle \phi (\bx) \rangle (t)|^2} 
  \int d \bx 
  \langle \phi^* (\bx) \rangle (t) \left( 
        -\frac{\hbar^2 \nabla^2}{2 m} + V^{\rm ext} (\bx)  + \frac{T^{\rm 2B}}{2} 
    | \langle \phi (\bx) \rangle (t)|^2 
  \right) \langle \phi (\bx) \rangle (t). 
\end{equation} 
 
We now show that the above `classical' approximation to the 
fluctuation-dissipation theorem indeed leads to the correct 
equilibrium. 
 Let us therefore substitute Eq.~(\ref{fdtheorem}) into the 
Fokker-Planck equation, which simplifies to 
\begin{eqnarray} 
\label{fpeqncondensatefdt} 
  && i \hbar \frac{\partial}{\partial t} P [\phi^*,\phi;t] = \nonumber \\ 
  &&-\frac{\beta}{4} \int d \bx \ \hbar \Sigma^{\rm K} \args \frac{\delta}{\delta \phi (\bx)} 
    \left( 
    - \frac{\hbar^2 \nabla^2}{2m} 
             + V^{\rm{ext}}(\bx) - \mu 
             + T^{\rm 2B} |\phi (\bx)|^2 
  \right) \phi (\bx) P [\phi^*,\phi;t] \nonumber \\ 
  && - \frac{\beta}{4} \int d \bx \ \hbar \Sigma^{\rm K} \args  \frac{\delta}{\delta \phi^* (\bx)} 
     \left( 
    - \frac{\hbar^2 \nabla^2}{2m} 
             + V^{\rm{ext}}(\bx) - \mu 
         + T^{\rm 2B} |\phi (\bx)|^2 
  \right) \phi^* (\bx) P [\phi^*,\phi;t]  \nonumber \\ 
  &&-\half \int d \bx \frac{\delta^2}{\delta \phi (\bx) \delta \phi^* (\bx)} 
         \hbar \Sigma^{\rm K} \args P[\phi^*,\phi;t]. 
\end{eqnarray} 
The stationary solution of this Fokker-Planck equation is given by 
\begin{equation} 
\label{equildistr} 
  P[\phi^*,\phi;t \to \infty] \propto \exp \left\{ -\beta 
    \int d \bx \ \phi^* (\bx) \left( 
        -\frac{\hbar^2 \nabla^2}{2 m} + V^{\rm ext} (\bx) -\mu + \frac{T^{\rm 2B}}{2} 
    | \phi (\bx) |^2 
  \right) \phi(\bx) \right\}, 
\end{equation} 
as can be checked by substitution.  To see that 
Eq.~(\ref{equildistr}) is in fact the correct equilibrium 
distribution, we have to show that the macroscopic condensate 
wave function $\langle \phi (\bx) \rangle$ obeys the 
time-independent Gross-Pitaevskii equation. To see this, we first 
note that 
\begin{equation} 
  \int d [\phi^*] d [\phi] \frac{\delta}{\delta \phi^* (\bx)} 
  P[\phi^*,\phi;t]=0 
\end{equation} 
for a general probability distribution which vanishes at the 
boundaries of the domain of integration. If we apply this to the 
equilibrium distribution $P[|\phi|;t \to \infty] $ we get, by applying 
the mean-field approximation $\langle |\phi (\bx)|^2 \phi (\bx) 
\rangle \simeq | \langle \phi (\bx) \rangle |^2 \langle \phi 
(\bx) \rangle $, the desired result 
\begin{equation} 
\label{gptimeindep} 
  \left( 
        -\frac{\hbar^2 \nabla^2}{2 m} + V^{\rm ext} (\bx) - \mu +T^{\rm 2B} 
    | \langle \phi (\bx) \rangle|^2 
  \right) \langle \phi (\bx) \rangle = 0, 
\end{equation} 
which is precisely the time-independent Gross-Pitaevskii equation. 
Note that Eq.~(\ref{fdtheorem}) together with the time-independent 
Gross-Pitaevskii equation implies that in 
equilibrium $\langle R (\bx,t) \rangle =0$. This means that 
there is detailed balance between the condensate and the thermal 
cloud, and that there is, on average, no condensate growth or 
evaporation when the system has relaxed to equilibrium. 
 
Using the results of the previous section, we now give a 
formulation of the nonequilibrium theory discussed above in terms 
of a Langevin field equation corresponding to the Fokker-Planck 
equation in Eq.~(\ref{fpeqncondensatefdt}). This Langevin field 
equation takes the form of a dissipative nonlinear Schr\"odinger 
equation with noise, given by 
\begin{eqnarray} 
\label{snlse} 
   i \hbar \frac{\partial \phi (\bx, t)}{\partial t}&=& 
   \left( 1+ \frac{\beta}{4} \hbar \Sigma^{\rm K} \args \right) \nonumber \\ 
     && \times \left\{ - \frac{\hbar^2 \nabla^2}{2m} 
             + V^{\rm{ext}}(\bx) - \mu 
         + T^{\rm 2B} |\phi \args|^2 
     \right\}  \phi \args + \eta \args. 
\end{eqnarray} 
This Langevin equation quite generally generalizes the Gross-Pitaevskii equation 
 to nonzero temperatures, and includes 
both dissipation and thermal fluctuations. The complex gaussian 
noise in the Langevin field equation has correlations 
\cite{henk2,michiel1} 
\begin{equation} 
\label{complexnoisecor} 
  \langle \eta^* \args \eta (\bx',t') \rangle = 
    \frac{i \hbar^2}{2}  \Sigma^{\rm K} \args 
      \delta (t-t') \delta (\bx-\bx'), 
\end{equation} 
where the strength of the noise is determined by a Keldysh self-energy, 
given by 
\begin{eqnarray} 
\label{sigmak} 
  \hbar \Sigma^{\rm K} \args & = & -4 \pi i \left( T^{\rm 2B}\right)^2 
                           \int \frac{d \bk_1}{(2 \pi)^3} 
                           \int \frac{d \bk_2}{(2 \pi)^3} 
               \int \frac{d \bk_3}{(2 \pi)^3} 
               (2 \pi)^3 \delta( \bk_1-\bk_2-\bk_3) 
                                  \nonumber\\ 
           &   & \times \delta \left( \langle \mu_{\rm c} \args \rangle +\epsilon_1 - 
                               \epsilon_2 -\epsilon_3 \right) 
                 \left[ N_1 (1+N_2) (1+N_3) + (1+N_1) N_2 N_3 \right]. 
\end{eqnarray} 
In this expression, $N_i$ is again the Bose-distribution function of the thermal 
cloud, evaluated at an energy of a thermal particle, which is in the 
Hartree-Fock approximation given by 
\begin{equation} 
\label{energythermparttimedep} 
  \epsilon_i = \frac{\hbar^2 \bk^2_i}{2 m} + V^{\rm{ext}} (\bx) 
              +  2 T^{\rm 2B} |\langle \phi (\bx,t) \rangle|^2. 
\end{equation} 
The average local chemical potential of the condensate atoms 
$\langle \mu_{\rm c} \args\rangle $, is given by 
Eq.~(\ref{chempotcondfunctional}). Notice that in the latter 
equation, and in the above expression for the energy of a thermal 
particle, $\langle \phi \args \rangle$ has to be determined 
self-consistently, since only then the probability distribution 
generated by the Langevin equation in Eq.~(\ref{snlse}) relaxes 
to the correct equilibrium. 
 
The stochastic non-linear Schr\"odinger equation in Eq.~(\ref{snlse}), 
together with the expression for the Keldysh self-energy in 
Eq.~(\ref{sigmak}), 
 gives a nonequilibrium description of the 
condensate, that obeys the fluctuation-dissipation theorem. In 
Sec. IV we use a variational {\it ansatz} to solve this equation. 
However, we first derive the corresponding noisy hydrodynamic 
formulation. 
 
\subsection{Stochastic hydrodynamics} 
The condensate is often described in terms of its density and its 
phase, by making the transformation $\phi = \sqrt{\rho} e^{i 
\theta}$. When applied to the Gross-Pitaevskii equation, this 
transformation results in the so-called Josephson equation for 
the phase, and a continuity equation for the density 
\cite{gps,zgn}. We now want to derive the generalization of these 
two equations to the case of our Langevin equation for the 
condensate. To do this, we first substitute the `classical' 
approximation to the fluctuation-dissipation theorem 
 into the effective action $S^{\rm{eff}} [\phi^*, \phi]$, which now reads 
\begin{eqnarray} 
\label{sefffd} 
  S^{\rm{eff}} [\phi^*, \phi] &=& \int_{t_0}^{t} dt' \int d \bx 
  \frac{2}{\hbar \Sigma^{\rm K} (\bx,t')} 
     \left| 
        \left( 
      i \hbar \frac{\partial}{\partial t'} + 
      \left\{ 1+\frac{\beta}{4} 
      \hbar\Sigma^{\rm K}  (\bx,t') \right\} \right. \right.  \nonumber \\ 
     && \ \ \ \ \times \left. \left. \left[ \frac{\hbar^2 \nabla^2}{2 m} 
      -V^{\rm ext} (\bx) + \mu 
      + T^{\rm 2B} |\phi (\bx,t')|^2 \right] 
    \right) \phi (\bx,t') 
     \right|^2. 
\end{eqnarray} 
The reason for this substitution is that we have defined the 
fluctuation-dissipation theorem as an operator equation in the 
configuration space of the order parameter, and not in terms of its density 
and phase. We can now easily substitute $\phi=\sqrt{\rho} e^{i \theta}$ 
into this effective action. This substitution results in an 
effective action in terms of the density and the phase of the order 
parameter, i.e., 
\begin{eqnarray} 
\label{seffrhotheta} 
  S^{\rm{eff}} [\rho,\theta] &=& \int_{t_0}^t dt' \int d \bx 
     \frac{2}{\hbar \Sigma^{\rm K} (\bx,t')}  \nonumber \\ 
     && \Biggl\{ \rho (\bx,t') 
      \Biggl( 
        \hbar \frac{\partial \theta (\bx,t')}{\partial t'} 
          - \frac{\beta}{4} i \hbar \Sigma^{\rm K} (\bx,t') 
            \frac{\hbar \mbox{\boldmath $\nabla$} \cdot (\rho (\bx,t') {\bf v}_{\rm s} (\bx,t'))}{2 \rho (\bx,t')} 
          \nonumber \\ 
        && \ \ \ \ \ \ \ \ \ \ \ \ \ \ \ \ 
            + \mu_{\rm c} (\bx,t')  - \mu 
        \Biggr)^2 \nonumber \\ 
        &+&  \frac{\hbar^2}{4 \rho (\bx,t')} 
          \left( 
           \frac{\partial \rho (\bx,t')}{\partial t'} 
            + \mbox{\boldmath $\nabla$} \cdot (\rho (\bx,t') {\bf v}_{\rm s} (\bx,t')) \right. \nonumber \\ 
       && \left. \ \ \ \ \ \ \ \ \ \ \ \ \ \ \ \ 
            +\frac{\beta}{2} i \Sigma^{\rm K} (\bx,t') 
             \left( \mu_{\rm c} (\bx,t')   - \mu 
             \right) \rho(\bx,t') 
          \right)^2 \Biggr\}, 
\end{eqnarray} 
where we used that $\Sigma^{\rm K} (\bx,t)$ only has a negative 
imaginary part, as seen from Eq.~(\ref{sigmak}), and thus $i 
\Sigma^{\rm K} (\bx,t)$ is a positive and real quantity. We 
defined the superfluid velocity ${\bf v}_{\rm s} \args$, and the 
condensate chemical potential $\mu_{\rm c} \args$ by means of 
\cite{voetnoot}, 
\begin{eqnarray} 
\label{defsvsmuc} 
  \mu_{\rm c} \args &=& - \frac{\hbar^2 \nabla^2 \sqrt{\rho \args}}{2 m \sqrt{\rho \args}} 
                + V^{\rm ext} (\bx) + T^{\rm 2B} \rho \args + \half m {\bf v}_{\rm s}^2 
        \args ; \nonumber \\ 
  {\bf v}_{\rm s} \args &=& \frac{\hbar}{m} 
                            \mbox{\boldmath $\nabla$}\theta \args , 
\end{eqnarray} 
which coincides with the expression in Eq.~(\ref{chempotcondfunctional}). 
The effective action $S^{\rm{eff}} [\rho,\theta]$ yields two stochastic 
equations of motion. The equation for the phase of the condensate takes the 
form of a stochastic Josephson equation 
\begin{equation} 
\label{stochjoseph} 
   \hbar \frac{\partial \theta \args}{\partial t} 
          - \frac{\beta}{4} i \hbar \Sigma^{\rm K} (\bx,t) 
            \frac{\hbar^2 \mbox{\boldmath $\nabla$} 
            \cdot (\rho \args \mbox{\boldmath $\nabla$}\theta \args)}{2m \rho 
        \args} = \mu - \mu_{\rm c} \args  + \frac{\nu 
        \args}{\sqrt{\rho \args}} . 
\end{equation} 
Here, the real gaussian noise $\nu \args$ has correlations given by 
\begin{equation} 
 \langle \nu \args \nu (\bx',t') \rangle = 
  \frac{i \hbar^2}{4} \hbar \Sigma^{\rm K} (\bx,t) \delta (t-t') \delta (\bx-\bx'). 
\end{equation} 
The stochastic Josephson equation has two modifications with 
respect to the ordinary Josephson equation. First, it has a 
spatial diffusion-like term proportional to $i \hbar \Sigma^{\rm 
K} (\bx,t)$. This term will cause the phase to undergo spatial 
diffusion due to collisions of thermal atoms with the condensate 
atoms, not to be confused with the phenomenon of phase 
`diffusion', which corresponds to spreading of the global phase 
due to quantum fluctuations \cite{liyou}, and therefore relax to 
a state where the phase is position independent. So, in 
equilibrium we have $\langle {\bf v}_{\rm s} \rangle \equiv \hbar 
\langle \mbox{\boldmath $\nabla$} \theta \rangle/m=0$, as 
expected. We will see lateron that this tendency towards 
equilibrium will give rise to an increase in the sound velocity 
in the Bose condensate. Secondly, the Josephson equation has a 
noise term inversely proportional to the square root of the 
density. This noise represents the fluctuations in the phase of 
the condensate due to incoherent collisions of thermal atoms with 
condensate atoms, i.e., 
 due to thermal fluctuations. 
 
The equation of motion for the density is a stochastic continuity 
equation with a source term, 
\begin{eqnarray} 
\label{continuityeqn} 
  \frac{\partial \rho \args}{\partial t} 
    &+& \mbox{\boldmath $\nabla$} \cdot (\rho \args {\bf v}_{\rm s} \args) \nonumber \\ 
    &&= -\frac{\beta}{2} i \Sigma^{\rm K} (\bx,t) 
             \left( \mu_{\rm c} \args  - \mu 
             \right) \rho \args + 2 \sqrt{\rho \args} \xi \args, 
\end{eqnarray} 
with correlations of the gaussian noise $\xi \args$ given by 
\begin{equation} 
\label{rhonoisecors} 
  \langle \xi \args \xi (\bx',t') \rangle 
    = \frac{i \Sigma^{\rm K} (\bx,t)}{4}  \delta (\bx-\bx') \delta (t-t'). 
\end{equation} 
In appendix A it is explained that we have to interpret this 
noise as a Stratonovich process. This gives rise to additional 
drift terms in the equation of motion for the average of the 
density, because in Eq.~(\ref{continuityeqn}) we are dealing with 
multiplicative noise. Note that from Eq.~(\ref{continuityeqn}) it 
is explicitly seen that there is condensate growth if 
\mbox{$\mu>\mu_{\rm c}$}, i.e., if the chemical potential lies 
above the chemical potential of the condensate. If 
\mbox{$\mu<\mu_{\rm c}$}, there is condensate evaporation. 
 
We will omit here the Fokker-Planck equation in terms of $\rho$ 
and $\theta$, but only discuss the equilibrium distribution 
generated by Eqs.~(\ref{stochjoseph}) and (\ref{continuityeqn}). 
It is simply determined from $P[\phi^*,\phi; t \to \infty]$ in 
Eq.~(\ref{equildistr}) by the substitution $\phi = \sqrt{\rho} 
e^{i \theta}$, since the jacobian of this transformation is equal 
to one. So we have 
\begin{eqnarray} 
  P[\rho,\theta;t \to \infty] \propto \exp \left\{ 
          -\beta 
         \int d \bx 
         \rho (\bx) 
         \left( 
           - \frac{\hbar^2 \nabla^2 \sqrt{\rho (\bx)}}{2 m \sqrt{\rho 
           (\bx)}} 
                + V^{\rm ext} (\bx) 
        + \frac{T^{\rm 2B}}{2} \rho (\bx) 
            + \half m {\bf v}_{\rm s}^2 (\bx) - \mu 
         \right) 
        \right\}. \nonumber 
\end{eqnarray} 
We see from this probability distribution that $\langle {\bf 
v}_{\rm s} \rangle=0$, which should be the case in equilibrium. 
The average density profile is again determined by the 
time-independent Gross-Pitaevskii equation, as explained before. 
 
To discuss the physical content of the stochastic continuity equation 
 and the stochastic Josephson equation 
further, we now derive the wave equation describing the 
propagation of sound waves in a Bose Einstein condensate. For 
simplicity, we discuss here the homogeneous case, where $V^{\rm 
ext} (\bx) =0$. A treatment of the trapped case is presented in 
Sec.V~A. We linearize the averages of Eqs.~(\ref{continuityeqn}) 
and (\ref{stochjoseph}) around their equilibrium solutions 
$\langle \rho \args \rangle=\rho_0$ and $\langle {\bf v}_{\rm s} 
\args \rangle=0$. Therefore, we write $\langle \rho \args 
\rangle=\rho_0 + \delta \rho \args$, and $\langle {\bf v}_s \args 
\rangle = \delta {\bf v}_{\rm s} \args$, and substitute this into 
the average of Eqs.~(\ref{stochjoseph}) and(\ref{continuityeqn}). 
Linearization results in two coupled equations of motion for the 
deviations, i.e., 
\begin{eqnarray} 
  m \frac{\partial \delta { \bf v}_s \args}{\partial t} && 
    = -T^{\rm 2B} \mbox{\boldmath $\nabla$} (\delta \rho \args) 
      +\frac{\beta i \hbar^2 \Sigma^{\rm K}}{8} \nabla^2 (\delta {\bf v}_{\rm s} 
      \args); \nonumber \\ 
  \frac{\partial \delta \rho \args}{\partial t} && 
    + \rho_0 \mbox{\boldmath $\nabla$} \cdot (\delta {\bf v}_{\rm s} \args) 
    = - \frac{\beta}{2} i \Sigma^{\rm K} \left( 2 T^{\rm 2B} \rho_0 
           - \mu \right) \delta \rho \args. 
\end{eqnarray} 
Note that we have made use of the fact that $\hbar \Sigma^{\rm K}$ is independent 
of the spatial coordinates for a homogeneous Bose gas, as can be seen from 
Eq.~(\ref{sigmak}). Next, we combine these two equations to obtain a single damped 
wave equation for the propagation of sound waves in a homogeneous Bose gas, 
\begin{equation} 
\label{waveeqn} 
  \left( \frac{\partial^2 }{\partial t^2}  - c^2 \nabla^2 \right) 
  \delta \rho \args= - \frac{1}{\tau} 
  \frac{\partial \delta \rho \args}{\partial t}. 
\end{equation} 
The relaxation time $\tau$ is defined to be inversely 
proportional to the damping rate of the waves. Physically, this 
damping arises because the excitation of a sound wave slightly 
disturbs the equilibrium situation where the average growth or 
evaporation of the condensate is equal to zero. Hence, there is 
no longer detailed balance between the condensate and the thermal 
cloud, and the collisions between the condensate and thermal 
atoms drive the condensate back to the equilibrium situation, 
where $ \delta  \rho \args =0$, and $ \delta {\bf v}_{\rm s} \args 
= 0$. The relaxation time $\tau$ is given by 
\begin{equation} 
\label{tau} 
  \frac{1}{\tau} = \frac{\beta}{2} i \Sigma^{\rm K}  T^{\rm 2B} \rho_0, 
\end{equation} 
where we used the time-independent Gross-Pitaevskii equation 
which reduces in this case to $\mu = T^{\rm 2B} \rho_0$, to 
eliminate the chemical potential. The sound velocity $c$ in 
Eq.~(\ref{waveeqn}) is given by 
\begin{equation} 
\label{soundvelo} 
  c^2=c_0^2 \left[ 1 + 
  \frac{1}{16} \left( \beta i \hbar \Sigma^{\rm K} \right)^2 
   \right], 
\end{equation} 
where $c_0=(T^{\rm 2B} \rho_0 /m)^{1/2}$ is the well-known zero 
temperature sound velocity, predicted by the Gross-Pitaevskii 
equation, and first obtained by Bogoliubov \cite{bogoliubov}. We 
see that our nonequilibrium treatment results in an increased 
sound velocity. This increase is a result from the term in the 
stochastic Josephson equation in Eq.~(\ref{stochjoseph}) 
proportional to $i \hbar \Sigma^{\rm K}$. Physically, this term 
represents the fact that the phase of the condensate undergoes 
spatial diffusion due to collisions between condensate and 
thermal atoms, and therefore relaxes to a state where $\langle 
{\bf v}_{\rm s} \rangle=0$. The spatial diffusion of the phase 
therefore increases the `stiffness' of the condensate, and hence 
results in an increase of the sound velocity. Since the increase 
in the sound velocity is of order ${\mathcal O} ( |\beta \hbar 
\Sigma^{\rm K}|^2 )$, its effect is in general small below the 
critical temperature, as the collisionless limit is determined by 
$|\beta \hbar \Sigma^{\rm K}| \ll 1$ and experiments are usually 
in this limit. The damped wave equation in Eq.~(\ref{waveeqn}) 
should be compared to the result found by Williams and Griffin 
\cite{williams1}. These authors use a dissipative non-linear 
Sch\"odinger equation, with a damping term similar to 
 Eq.~(\ref{rterm}), to arrive at a damped wave equation describing the propagation 
of sound in a trapped Bose-Einstein condensate in the presence of 
a static thermal cloud. Note, however, that although the 
microscopic expression used by Williams and Griffin is of the 
same form as in Eq.~(\ref{rterm}), the chemical potential of the 
condensate used by these authors in the calculation of $R \args$ 
is not the operator given by Eq.~(\ref{hamgp}). Instead, they use 
for the condensate energy in the energy-conserving delta function 
the expression $\langle \mu_{\rm c} \args \rangle$, which in 
principle violates the fluctuation-dissipation theorem. As a 
consequence, these authors do not find an increase in the sound 
velocity at nonzero temperatures. 
 
\section{Variational approximation} 
Although the stochastic non-linear Schr\"odinger equation given in 
Eq.~(\ref{snlse}), or equivalently, the hydrodynamic formulation 
given in Eqs.~(\ref{stochjoseph}) and (\ref{continuityeqn}), give 
a full nonequilibrium description of the condensate that can in 
principle be solved numerically \cite{michiel1}, we find it more 
convenient to make analytical progress. Therefore, in the case of 
a harmonic trapping potential $V^{\rm ext} (\bx )= \sum_j m 
\omega_j x^2_j/2$, we consider a gaussian variational {\it 
ansatz} for the condensate wave function 
\begin{equation} 
\label{ansatz} 
  \phi \args = \sqrt{N_{\rm c} (t)} e^{i \theta_0 (t)} \prod_j \left( \frac{1}{\pi 
  q_j^2(t)} \right)^{\frac{1}{4}} \exp \left\{ -\frac{x_j^2}{2 q_j^2 (t)} \left( 1- 
  \frac{i m}{\hbar} q_j(t) \dot q_j (t) \right) \right\}. 
\end{equation} 
Here, the variational parameters $q_j (t)$ denote the gaussian 
widths of the condensate in the three spatial directions. The 
wave function is normalized to the number of atoms in the 
condensate $N_{\rm c} (t)$. This {\it ansatz} is different from 
the ones used in previous work, in the sense that it also 
 contains a global phase $\theta_0 (t)$. This turns out to be 
crucial, since the number of particles $N_{\rm c} (t)$, which is 
the variable conjugate to the global phase, is not constant in 
our case. Therefore, one must also allow for fluctuations in the 
global phase $\theta_0 (t)$ of the condensate. We expect this 
{\it ansatz} to give correct results when the number of particles 
is small, because the mean-field interaction of the condensate 
will then be small, and the condensate density profile will be 
close to the ideal gas solution. Moreover, it has also proven to 
give correct results for the frequencies of the collective modes 
of the condensate, even in the Thomas-Fermi regime, where the 
mean-field interaction, and thus the number of atoms in the 
condensate, is large \cite{gauss}. Therefore, we expect also to 
obtain physically sensible results even in this case. 
 
When the gaussian {\it ansatz} is applied to the Gross-Pitaevskii equation 
 we find that the variational parameters $q_j (t)$ obey 
Newton's equations of motion \cite{henk3,gauss,legget} 
\begin{equation} 
\label{classeqns} 
  \half m N_{\rm c} (t) \ddot q_j (t) = - 
   \frac{\partial V}{\partial q_j} ({\bf q} (t), N_{\rm c} (t)), 
\end{equation} 
with a potential energy equal to 
\begin{equation} 
\label{potential} 
   V ({\bf q},N_{\rm c}) = \sum_j \left( \frac{N_{\rm c} \hbar^2}{4 m q_j^2} + \frac{1}{4} m 
  N_{\rm c} \omega_j^2 q_j^2 
                       \right) 
  + \frac{a \hbar^2 N_{\rm c}^2}{\sqrt{2 \pi} m q_x q_y q_z}. 
\end{equation} 
From the action in Eq.~(\ref{sefffd}), we want to derive similar 
equations of motion, extended to the nonequilibrium case. For 
simplicity, we first consider an ideal gas, i.e., we drop the 
mean-field interaction term $T^{\rm 2B} |\phi \args|^2$. The 
condensate remains, however, in contact with the thermal cloud, 
that acts as a `heat bath'. Secondly, we assume that the Keldysh 
self-energy is constant over the size of the condensate. Although 
this assumption is not justified in general, we can always 
approximately compensate for this, by calculating a position 
independent Keldysh self-energy $\hbar \Sigma^{\rm K} (t)$ by 
means of an 
 appropriately averaged $\hbar \Sigma^{\rm K} \args$ over the size of the condensate. 
 We substitute our trial wave function 
 into the effective action in Eq.~(\ref{sefffd}), to obtain a probability 
distribution in terms of $N_{\rm c}, \theta_0$ and ${\bf q}$. It 
is given by 
\begin{equation} 
  P[N_{\rm c},\theta_0,{\bf q};t] = \int 
  d[N_{\rm c}] \ d[\theta_0]  \ d[{\bf q}] 
    \exp \left\{ \frac{i}{\hbar} S^{\rm eff} [N_{\rm c},\theta_0,{\bf q}] 
    \right\}, 
\end{equation} 
with an effective action that reads 
\begin{eqnarray} 
\label{seffq} 
  S^{\rm{eff}} [N_{\rm c},\theta_0;{\bf q}] &=& 
    \int_{t_0}^t dt'  \frac{2}{\hbar \Sigma^{\rm K} (t')}  \Biggl\{ 
      N_{\rm c} (t') \left( 
         \hbar \frac{d \theta_0 (t')}{d t'} 
         + \mu_{\rm c} (t')-\sum_j \frac{1}{4} m \dot q_j^2 (t') 
         + \sum_j \frac{1}{4} m q_j (t') \ddot q_j (t') 
         -\mu 
        \right)^2  \nonumber \\ 
   && + \frac{\hbar^2}{4 N_{\rm c} (t')} \left( 
        \frac{d N_{\rm c} (t')}{d t'} 
         + \frac{\beta}{2} i \Sigma^{\rm K} (t') 
           \left[ \mu_{\rm c} (t') 
              - \mu 
           \right] 
        N_{\rm c} (t') 
        \right)^2 \nonumber \\ 
   && + \sum_j \frac{q_j^2 (t')}{2 N_{\rm c}(t')} 
        \left( 
          \frac{1}{2} m N_{\rm c} (t') \ddot q_j(t') 
          + \frac{N_{\rm c} (t') \beta}{4} i \hbar^2 \Sigma^{\rm K} (t') 
            \frac{\dot q_j (t')}{q_j^2 (t')} 
          + \frac{\partial V}{\partial q_j} ({\bf q} (t'),N_{\rm c} (t')) 
        \right)^2 \nonumber \\ 
   && +  {\mathcal O} \left( (\beta i \hbar \Sigma^{\rm K})^2 \right) 
          \Biggr\}. 
\end{eqnarray} 
The potential in this effective action is defined by 
\begin{equation} 
\label{nonintpotential} 
  V ({\bf q},N_{\rm c}) = \sum_j \left( \frac{N_{\rm c} \hbar^2}{4 m q_j^2} + \frac{1}{4} m 
  N_{\rm c} \omega_j^2 q_j^2 
                       \right), 
\end{equation} 
which is precisely the potential given by Eq.~(\ref{potential}), without the 
mean-field interaction term. The condensate chemical potential for the 
gaussian {\it ansatz} is given by 
\begin{equation} 
\label{chempotgauss} 
  \mu_{\rm c} (t) = \frac{\partial V}{\partial N_{\rm c}} ({\bf q}(t),N_{\rm c}(t)) + 
    \sum_j \frac{1}{4} m \dot q_j^2 (t), 
\end{equation} 
as expected. We now assume the dimensionless parameter $\beta i 
\hbar \Sigma^{\rm K}$ to be small, and thus restrict ourselves to 
a temperature regime sufficiently far below the critical 
temperature, where $| \beta \hbar \Sigma^{\rm K} | \ll 1$, i.e., 
the collisionless regime. We can then to a good approximation 
neglect the terms quadratic in $\beta i \hbar \Sigma^{\rm K}$. 
The effective action in Eq.~(\ref{seffq}) thus becomes a sum of 
three squares, and we can extract equations of motion with 
gaussian noise terms, exactly as in Sec. II. Since the action is 
quadratic in $\theta_0 (t)$, we can integrate over this global 
phase exactly, because it only requires a gaussian integral. 
However, before we perform this integration, we discuss the 
stochastic equation of motion for $\theta_0 (t)$. With the 
techniques discussed in the Sec. II, we easily see that it is 
given by 
\begin{equation} 
\label{eqntheta0} 
  \hbar \frac{d \theta_0 (t)}{d t}  = \mu - 
   \mu_{\rm c}(t) + \sum_j \frac{1}{4} m \dot q_j^2 (t) 
         - \sum_j \frac{1}{4} m q_j (t) \ddot q_j (t) 
  + \frac{\nu (t)}{\sqrt{N_{\rm c} (t)}}, 
\end{equation} 
with time correlation of the noise 
\begin{equation} 
\label{thetanoise} 
\langle \nu (t') \nu (t) \rangle = 
    \frac{i \hbar^2 \Sigma^{\rm K} (t)}{4} \delta (t' -t). 
\end{equation} 
This stochastic equation again has the form of a Josephson 
equation with a noise term added, similar to 
Eq.~(\ref{stochjoseph}). Note that the noise term in 
Eq.~(\ref{eqntheta0}) is inversely proportional to the square 
root of the number of particles in the condensate. As a result, 
the Fokker-Planck equation for the probability distribution of 
$\theta_0 (t)$, associated with the Langevin equation in 
Eq.~(\ref{eqntheta0}), will have a `diffusion' term inversely 
proportional to the number of particles. This means that the 
global phase is only well determined if there is a infinite number 
of atoms in the condensate, otherwise the global phase undergoes 
phase diffusion, due to thermal fluctuations. This mechanism for 
phase diffusion is different than the phase `diffusion' 
considered by Lewenstein and You \cite{liyou}, who considered 
phase spreading due to quantum fluctuations. 
 
Having made these remarks, we perform the integration over 
$\theta_0 (t)$, and are left with a probability distribution for 
$N_{\rm c}$ and ${\bf q}$. It is given by 
\begin{equation} 
\label{probdistrnq} 
 P[N_{\rm c},{\bf q},{\bf v};t] = 
 \int d[N_{\rm c}] \ d[{\bf 
 q}] \ d[{\bf v}] \ \delta [{\bf v} (t')-\dot {\bf q}(t')] \ 
 \exp \left\{ \frac{i}{\hbar} S^{\rm eff} [N_{\rm c},{\bf q,{\bf v}}] \right\}. 
\end{equation} 
Here we introduced the velocity ${\bf v}(t)=\dot {\bf q}(t)$ by means of a 
delta functional. The resulting effective action reads 
\begin{eqnarray} 
\label{seffqv} 
  S^{\rm{eff}} [N_{\rm c},{\bf q},{\bf v}] &=& 
    \int_{t_0}^t dt'  \frac{2}{\hbar \Sigma^{\rm K} (t')}  \Biggl\{ 
     \frac{\hbar^2}{4 N_{\rm c} (t')} \left( 
        \frac{d N_{\rm c} (t')}{d t'} 
         + \frac{\beta}{2} i \Sigma^{\rm K} (t') 
           \left[ \mu_{\rm c} (t') 
                  - \mu 
           \right] 
        N_{\rm c} (t') 
        \right)^2 \nonumber \\ 
   &+& \sum_j \frac{q_j^2 (t')}{2 N_{\rm c}(t')} 
        \left( 
          \frac{1}{2} m N_{\rm c} (t') \dot v_j (t') 
          + \frac{N_{\rm c} (t') \beta}{4} i \hbar^2 \Sigma^{\rm K} (t') 
            \frac{v_j (t')}{q_j^2 (t')} 
          + \frac{\partial V}{\partial q_j} ({\bf q} (t'),N_{\rm c} (t')) 
        \right)^2 \Biggr\} 
\end{eqnarray} 
Eqs.~(\ref{probdistrnq}) and (\ref{seffqv}) are similar to the 
path-integral expressions we encountered in our discussion of the 
Brownian motion of a particle in a potential in Sec. II. Therefore we immediately 
conclude that the equations of motion for the variational parameters are 
given by 
\begin{equation} 
\label{qlangevin} 
   \frac{1}{2} m N_{\rm c} (t) \ddot q_j(t) 
          + \frac{N_{\rm c} (t) \beta}{4} i \hbar^2 \Sigma^{\rm K} (t) 
            \frac{\dot q_j (t)}{q_j^2 (t)} 
          =- \frac{\partial V}{\partial q_j} ({\bf q} (t),N_{\rm c} (t)) 
        + \frac{\sqrt{2 N_{\rm c} (t)}}{q_j(t)} \xi_j (t), 
\end{equation} 
with the time correlations of the gaussian noise terms $\xi_j (t)$ given by 
\begin{equation} 
\label{qnoisetimecor} 
  \langle \xi_j (t) \xi_k (t') \rangle = \frac{i \hbar^2 \Sigma^{\rm K} (t)}{4} 
  \delta_{jk} \delta (t-t'). 
\end{equation} 
So we have found the important result that the variational 
parameters obey the equations of motion of a Brownian particle 
with mass $m N_{\rm c}/2$ in a potential $V ({\bf q},N_{\rm c})$. 
Physically, the variational description of the condensate with 
the Langevin equation in Eq.~(\ref{qlangevin}), as opposed to 
Eq.~(\ref{classeqns}), has two important extra features. First, 
there is a damping term present, i.e., a term proportional to the 
velocity $\dot q_j (t)$. This damping term can, for example, be 
used to calculate the damping on the collective modes of the 
condensate. Since the damping term is proportional to $\hbar 
\Sigma^{\rm K}$, we conclude that it arises because of incoherent 
collisions between condensate and thermal atoms, which drive the 
condensate back to equilibrium, and let the phase of the 
condensate relax to a state where the phase is, on average, 
position independent. Secondly, since the Langevin equation also 
contains fluctuations, it can, for example, be used to describe 
the stochastic initiation of the collapse observed in $^7$Li 
\cite{curtis,cass1,cass2,randy}. Our description contains thermal 
fluctuations, that cause the condensate to overcome the 
macroscopic energy barrier and start the collapse. In the next 
section, we will present the result of calculations that we have 
done on the two above mentioned phenomena. 
 
Since the potential in the Langevin equation in 
Eq.~(\ref{qlangevin}) depends on the number of condensate 
particles $N_{\rm c}$, we have to couple Eq.~(\ref{qlangevin}) to 
a rate equation for the number of condensate atoms. This 
stochastic rate equation also follows directly from the effective 
action in Eq.~(\ref{seffqv}) with the techniques discussed 
previously, and is given by 
\begin{eqnarray} 
\label{rateeqn} 
  \frac{d N_{\rm c} (t)}{d t} 
         =- \frac{\beta}{2} i \Sigma^{\rm K} (t) 
           \left[ \mu_{\rm c} (t) 
                  - \mu 
           \right] 
        N_{\rm c} (t) + 2 \sqrt{N_{\rm c} (t)} \eta (t), 
\end{eqnarray} 
with correlations of the gaussian noise given by 
\begin{equation} 
  \langle \eta (t') \eta (t) \rangle = 
    \frac{i \Sigma^{\rm K} (t)}{4} \delta (t' -t). 
\end{equation} 
As explained in appendix A, we have to treat the multiplicative 
noise in Eq.~(\ref{rateeqn}) again as a Stratonovich process, to 
achieve the correct equilibrium distribution. Physically, 
Eq.~(\ref{rateeqn}) describes the growth or evaporation of the 
condensate. The noise term in Eq.~(\ref{rateeqn}) represents the 
fluctuations in the number of particles. 
 
To see that Eqs.~(\ref{qlangevin}) and (\ref{rateeqn}) generate 
the correct equilibrium distribution, we now discuss the 
Fokker-Planck equation for $P[N_{\rm c},{\bf q}, {\bf v};t]$. 
However, let us first discuss the equilibrium solution we expect 
on basis of Eq.~(\ref{equildistr}). A substitution of the 
gaussian {\it ansatz} in Eq.~(\ref{ansatz}) into equation 
Eq.~(\ref{equildistr}) results in 
\begin{equation} 
\label{equildistrnqv} 
  P[N_{\rm c},{\bf q}, {\bf v}; t \to \infty] \propto 
    \exp 
     \left\{ 
       - \beta 
          \left( 
        \sum_j \frac{1}{4} m N_{\rm c} v_j^2 + V ({\bf q},N_{\rm c}) - \mu N_{\rm c} 
      \right) 
     \right\}, 
\end{equation} 
where $V ({\bf q},N_{\rm c})$ is the potential given by 
Eq.~(\ref{potential}). Although the Langevin equations for $q_j 
(t)$ and $N_{\rm c} (t)$ did, in first instance, not include the 
mean-field interactions, we argue that they also are correct for 
the interacting case. The reason for this is, that in this manner 
we are led to the correct equilibrium distribution as we show 
now. Let us therefore determine the Fokker-Planck equation for the 
probability distribution of $N_{\rm c}$, ${\bf q}$ and ${\bf v}$, 
generated by the stochastic equations in Eqs.~(\ref{qlangevin}) 
and (\ref{rateeqn}) with the interacting potential in 
Eq.~(\ref{potential}). It is given by 
\begin{eqnarray} 
\label{fpeqnnqv} 
 && \frac{\partial P [N_{\rm c},{\bf q}, {\bf v};t]}{\partial t} = \nonumber \\ 
 && \sum_j \left[ 
       -\frac{\partial}{\partial q_j} v_j 
       +\frac{\partial}{\partial v_j} 
          \left( 
        \frac{\beta i \hbar^2 \Sigma^{\rm K} (t)}{2 m q_j^2}  v_j 
        +\frac{2}{m N_{\rm c}} \frac{\partial V}{\partial q_j} ({\bf q},N_{\rm c}) 
      \right) 
        + \frac{i \hbar^2 \Sigma^{\rm K} (t)}{m^2 N_{\rm c} q_j^2} 
      \frac{\partial^2}{\partial v_j^2} 
    \right] P[N_{\rm c},{\bf q}, {\bf v};t] + \nonumber \\ 
 && \left[ 
      \frac{\partial}{\partial N_{\rm c}} 
      \left( 
        \frac{\beta}{2} i \Sigma^{\rm K} (t) 
      \left[ 
        \frac{\partial V}{\partial N_{\rm c}} ({\bf q},N_{\rm c}) 
        +\sum_j \frac{1}{4} m v_j^2 - \mu 
      \right] N_{\rm c} 
      \right) 
      + \frac{i \Sigma^{\rm K}(t)}{2} 
        \frac{\partial}{\partial N_{\rm c}} N_{\rm c} 
    \frac{\partial}{\partial N_{\rm c}} 
    \right] P[N_{\rm c},{\bf q}, {\bf v};t], 
\end{eqnarray} 
and insertion of the equilibrium distribution shows that it is 
indeed a stationary solution of this Fokker-Planck equation. Thus 
we conclude that the Langevin equations for $q_j (t)$ and $N_{\rm 
c}(t)$ give, with the potential $V({\bf q},N_{\rm c})$ given in 
Eq.~(\ref{potential}), the correct description of the 
nonequilibrium dynamics of a Bose-Einstein condensate in the 
gaussian approximation. This description includes damping of the 
collective modes of the condensate, as well as condensate growth 
and evaporation. The essence of our method lies in the 
fluctuation-dissipation theorem, which ensures the relaxation 
towards the correct physical equilibrium distribution. 
 
\section{Applications} 
In this section we first apply the Langevin equations for the 
variational parameters, derived in the previous section, to the 
calculation of the damping and frequency of the collective modes 
of the condensate. As a second application, we also obtain a 
description of the initial growth of a condensate. 
\subsection{Collective modes of the condensate} 
In this section we use the Langevin equations in Eq.~(\ref{qlangevin}) for the 
gaussian variational parameters, and the stochastic rate equation 
in Eq.~(\ref{rateeqn}) for the number of particles in the condensate, 
 to obtain a description of the collective modes of the 
condensate. We calculate the frequency and damping of both the 
monopole and quadrupole mode in an isotropic trap, and compare 
with the theoretical results found by Williams and Griffin 
\cite{williams1,williams2}. Since we are considering the case of 
a static thermal cloud, our results will be correct only for the 
modes where the thermal cloud does not play an important role, 
i.e., for the out-of-phase modes \cite{michiel3,usama}. In the 
experiments of Jin {\it et al.} \cite{jin}, this turns out to be 
the quadrupole mode. We calculate the frequency of the quadrupole 
mode for this experiment, by means of a fit to the experimental 
data for the damping. 
 
The frequency and damping of the collective modes are, as 
measured in experiment, averaged 
quantities. Therefore, we first write down the equations of motion for the 
averages of the gaussian variational parameters and the number of particles 
in the condensate. The equations of motion for the average of the gaussian 
widths read 
\begin{equation} 
\label{qaverage} 
   \frac{1}{2} m N_{\rm c} (t) \ddot q_j(t) 
          + \frac{N_{\rm c} (t) \beta}{4} i \hbar^2 \Sigma^{\rm K} 
            \frac{\dot q_j (t)}{q_j^2 (t)} 
          =- \frac{\partial V}{\partial q_j} ({\bf q} (t),N_{\rm c} (t)). 
\end{equation} 
For notational convenience we omit the brackets $\langle \cdots 
\rangle$ denoting the noise average of a stochastic quantity, and 
denoted the averages of the gaussian variational parameters 
simply by $q_j (t)$, where $j$ equals $x$, $y$ or $z$. The average 
equation in Eq.~(\ref{qaverage}) is obtained from the Langevin 
equation for $q_j (t)$ by simply leaving out the noise term. This 
can be done because the noise in the Langevin equation does not 
induce a drift term for the average. This follows directly from 
the Fokker-Planck equation, with the use of partial integration. 
Since we want to describe a perturbation around a static 
equilibrium, the Keldysh self-energy will be time independent to a 
good approximation, and we thus drop its explicit dependence on 
time. For a description of the collective modes, we also have to 
consider variations in the average number of particles of the 
condensate, caused by the excitation of a mode. This means that 
we also have to consider the rate equation for the average number 
of particles 
\begin{equation} 
\label{averagerateeqn} 
  \frac{d N_{\rm c} (t)}{d t} 
         =- \frac{\beta}{2} i \Sigma^{\rm K} 
           \left[ \mu_{\rm c} (t) 
                  - \mu 
           \right] 
        N_{\rm c} (t) + \frac{i \Sigma^{\rm K}}{2}. 
\end{equation} 
In writing down this equation, we again left out the brackets 
$\langle \cdots \rangle$, which denote averaging over different 
realizations of the noise in the stochastic rate equation in 
Eq.~(\ref{rateeqn}). The last term on the right hand side of 
Eq.~(\ref{averagerateeqn}) is a so-called noise-induced, or 
spurious drift term. It arises because in Eq.~(\ref{rateeqn}) we 
are dealing with multiplicative Stratonovich noise. Without this 
drift term, the equilibrium number of particles predicted by 
Eq.~(\ref{averagerateeqn}) would not be correct, as we will see 
lateron. Note that the average of the stochastic rate equation in 
Eq.~(\ref{averagerateeqn}) is very similar to the result obtained 
by Gardiner {\it et al.} \cite{growthformula}. However, their 
expression for the chemical potential of the condensate is 
different since they do not consider a gaussian {\it ansatz}, and 
they also have not made the `classical' approximation to the 
fluctuation-dissipation theorem. 
 
To obtain a description of the collective modes of the 
condensate, we have to linearize the equations in 
Eqs.~(\ref{qaverage}) and (\ref{averagerateeqn}) around their 
time-independent equilibrium solutions. Let us therefore put 
$q_j(t)=q^{(0)}_j + \delta q_j (t)$ and $N_{\rm c}(t)=N_0+\delta 
N(t)$ and substitute this in Eqs.~(\ref{qaverage}) and 
(\ref{averagerateeqn}). Equating the zeroth-order terms after 
linearization results for the average rate equation 
 in 
\begin{equation} 
\label{nzeroth} 
  \left[ \beta \left( \frac{\partial V ({\bf q}^{(0)},N_0)}{\partial N_{\rm c}} -\mu 
  \right) - \frac{1}{N_0} \right] N_0 = 0. 
\end{equation} 
From this equation the equilibrium number of particles can be calculated. 
It is however much more convenient to use the number of particles in the 
condensate as experimental input, and calculate the chemical potential 
$\mu$ such that Eq.~(\ref{nzeroth}) is satisfied. 
The equilibrium conditions for the gaussian variational parameters read 
\begin{equation} 
\label{qzeroth} 
  \frac{\partial V ({\bf q}^{(0)},N_0)}{\partial q_j} = 0, 
\end{equation} 
from which ${\bf q}^{(0)}$ can be calculated. For the 
noninteracting 
 case, where $a=0$ in the potential, 
the above equations result in $q_j=\sqrt{\hbar/m 
\omega_j}$ as expected. In this case $N_0$ is given by 
\begin{equation} 
  N_0 = \left[ \beta (\frac{\hbar}{2} (\omega_x+\omega_y+\omega_z) - \mu) \right]^{-1}, 
\end{equation} 
which is the correct equilibrium ground state occupation 
number of a non-interacting Bose gas, within the `classical' approximation. 
Note that without the noise-induced drift term in 
the rate equation for the average number of particles 
the correct equilibrium would not have been obtained. 
 
The linearized equations of motion for the deviations are found by equating 
the first order contribution on the left and right-hand side of Eqs.~(\ref{qaverage}) 
and (\ref{averagerateeqn}) after linearization. The 
equation for the deviation in the equilibrium number $N_0$ of condensate 
atoms due to the excitation of a collective mode is given by 
\begin{equation} 
\label{nfirst} 
  \delta \dot N (t) = - \Gamma \ \delta N (t) -  \sum_j \alpha_j 
  \delta q_j (t). 
\end{equation} 
Here, the parameter $\Gamma$ is given by 
\begin{equation} 
\label{R} 
  \Gamma = -\frac{\beta}{2} i \Sigma^{\rm K} \left[ \frac{1}{\beta N_0} + N_0 
      \frac{\partial^2 V({\bf q}^{(0)},N_0)}{\partial N_{\rm c}^2} \right], 
\end{equation} 
where we eliminated the chemical potential $\mu$ by using Eq.~(\ref{nzeroth}). 
Physically, $\Gamma$ describes the lack of detailed balance 
between the thermal cloud and the condensate due to an excitation of a 
collective mode of the condensate. In general, this lack of detailed 
balance will cause damping, and will alter the frequency with respect to 
the undamped case. The parameters $\alpha_j$ are given by 
\begin{equation} 
\label{alphaj} 
  \alpha_j = \frac{\beta}{2} i \Sigma^{\rm K} \frac{\partial^2 V({\bf 
  q}^{(0)},N_0)}{\partial q_j \partial N_{\rm c}}, 
\end{equation} 
and represent the response of the fluctuations in the 
number of particles due to a deformation of the condensate in the $j$-th 
direction, and vice-versa. 
 
The linearized equations of motion for the deviation of the gaussian variational 
parameters take the form of damped harmonic equations 
\begin{equation} 
\label{qfirst} 
  \delta \ddot q_j (t) + \Gamma_j \delta \dot q_j (t) = - \alpha_j/\alpha 
   \ \delta N (t) - \Omega^2_j \delta q_j (t) -\sum_{k \not= j} \Omega_{jk}^2 \delta q_k (t). 
\end{equation} 
The damping rates $\Gamma_j$ are given by 
\begin{equation} 
\label{dampingj} 
  \Gamma_j = \frac{\beta}{2 m (q_j^{(0)})^2} i \hbar^2 \Sigma^{\rm K}, 
\end{equation} 
and the frequencies $\Omega_j$ and $\Omega_{jk}$ read 
\begin{equation} 
\label{qfreq} 
  \Omega_j^2 = 
    \frac{2}{m N_0} \frac{\partial^2 V({\bf 
        q}^{(0)},N_0)}{\partial q_j^2}; \ \ 
  \Omega_{jk}^2 = \frac{2}{m N_0} \frac{\partial^2 V({\bf 
        q}^{(0)},N_0)}{\partial q_j \partial q_k}. 
\end{equation} 
In Eq.~(\ref{qfirst}) we introduced the parameter $\alpha=m N_0 
\beta i \Sigma^{\rm K}/4$ for later convenience. Physically, the 
damping rates $\Gamma_j$ arise because of collisions between 
thermal atoms and condensate atoms. This causes damping of the 
collective modes on the condensate, and also alters the 
frequencies with respect to the results obtained without damping. 
It should be noted here, that all the parameters mentioned above 
can be calculated microscopically, by using the expression for 
the Keldysh self-energy given in Eq.~(\ref{sigmak}). 
 
To obtain the eigenmodes of Eqs.~(\ref{nfirst}) and (\ref{qfirst}) we 
have to consider solutions of the form $(\delta N (t), \delta {\bf q} (t)) 
= (\delta N (0),\delta {\bf q} (0)) e^{-i \omega t}$. For such solutions we can 
rewrite these equations as 
\begin{equation} 
\label{matrixeqn} 
  \left( 
    \begin{array}{cccc} 
       \alpha (\Gamma - i \omega)  & \alpha_x & \alpha_y & \alpha_z \\ 
      \alpha_x & \Omega_x^2-i \omega \Gamma_x - \omega^2 & 
      \Omega_{xy}^2 & \Omega_{xz}^2 \\ 
      \alpha_y & \Omega_{xy}^2 & \omega_{y}^2- i \omega \Gamma_y - 
      \omega^2 & \Omega_{yz}^2 \\ 
      \alpha_z & \Omega_{xz}^2 & \Omega_{yz}^2 &\Omega_z^2-i \omega 
      \Gamma_z - \omega^2 
    \end{array} 
  \right) 
  \left( 
    \begin{array}{c} 
     \delta N (0)/\alpha \\ 
     \delta q_x (0) \\ 
     \delta q_y (0) \\ 
     \delta q_z (0) 
    \end{array} 
  \right) = 0. 
\end{equation} 
This matrix equation only has non-trivial solutions if the determinant of 
the above matrix is equal to zero. Solving this condition for $\omega$ 
results in general in complex frequencies $\omega = \omega_{\rm R} - i 
\omega_{\rm I}$. 
The real part $\omega_{\rm R}$ then gives the frequency of a collective 
mode, whereas the imaginary part $\omega_{\rm I}$ gives the damping of this mode. 
 
\subsubsection{Isotropic trapping potential} 
We consider now the case of an isotropic trapping potential $V^{\rm ext} 
(\bx)= 
\half m \omega_0^2 \bx^2$ for a discussion of the frequencies and 
damping rate of the low-lying collective modes of the condensate. Because 
of the spherical symmetry of the condensate in equilibrium, 
the number of parameters reduces and the 
eigenvalue equation simplifies significantly. 
For the frequencies and damping rates we have: 
\begin{eqnarray} 
  \Omega_x = \Omega_y = \Omega_z &\equiv& \Omega_r; \nonumber \\ 
  \Omega_{xy} = \Omega_{xz} = \Omega_{yz} &\equiv& \Omega_{rr}; \nonumber \\ 
  \Gamma_x = \Gamma_y = \Gamma_z &\equiv& \Gamma_r . 
\end{eqnarray} 
For the parameters $\alpha_j$ we have the same simplification, 
and we denote these parameters by $\alpha_r$. With these 
simplifications, the eigenmodes can be calculated analytically. 
One mode is doubly degenerate, with eigenvectors $(0,1,-1,0)$ and 
$(0,1,0,-1)$, and is thus the quadrupole mode. From the form of 
the eigenvectors of the quadrupole mode, we are led to the 
important conclusion that the number of particles in the 
condensate is constant for this mode. Physically this can be 
understood from the fact that the motion of the condensate is 
`volume-preserving' in this case: as one direction shrinks the 
other one expands and vice versa. This motion does not lead to a 
change in the chemical potential of the condensate, at least to a 
linear approximation, and therefore does not affect the average 
number of atoms in the condensate. The complex frequency of the 
quadrupole mode is given by 
\begin{equation} 
\label{quadfreq} 
  \omega_{\rm quad} = 
  \sqrt{\Omega_r^2-\Omega_{rr}^2-\left( \frac{\Gamma_r}{2} \right)^2} 
  -\frac{1}{2} i \Gamma_r. 
\end{equation} 
Note that from this expression it is clearly seen that the damping also 
affects the frequencies of the collective modes \cite{usama}. 
 
The frequency and eigenvector of the monopole mode can also be 
calculated analytically for the isotropic case. However, because 
of the rather formidable expressions involved, we omit them here. 
For the monopole mode the number of atoms in the condensate is not 
constant, but oscillates out of phase with the spatial degrees of 
freedom of the condensate. This can be understood from the fact 
that the monopole motion leads to a global increase in the 
density of the condensate, and therefore affects the detailed 
balance of the condensate with the thermal cloud. In the case 
where we ignore the fluctuations of the number of atoms in the 
condensate, and take $\alpha=\alpha_r=\Gamma=0$, the expression 
for the complex frequency of the monopole mode is given by 
\begin{equation} 
\label{monofreq} 
  \omega_{\rm mono} = \sqrt{\Omega_r^2+2 \Omega_{rr}^2 - \left( \frac{\Gamma_r}{2} 
          \right)^2}-\frac{1}{2} i \Gamma_r. 
\end{equation} 
Comparing the results in Eqs.~(\ref{quadfreq}) and 
(\ref{monofreq}) we see that the damping rate of both modes is 
equal in first order in $\Gamma_r$, at least within our 
variational approximation. 
 
We now turn to an explicit calculation of the frequencies and 
damping rates of the quadrupole and monopole mode in an isotropic 
trap. We have used the same parameters as Williams and Griffin 
\cite{williams1,williams2}, and thus have taken the trapping 
potential frequency equal to $\omega_0/2 \pi = 10$ Hz. The 
calculations are performed for $^{87}$Rb, which has a scattering 
length of $a=5.7$ nm. We take the total number of atoms equal to 
$N_{\rm total}=2 \times 10^6$. Since the number of atoms in the 
condensate is large at most temperatures below the critical 
temperature, we are mainly in the Thomas-Fermi regime, and can 
neglect the kinetic energy of the condensate atoms with respect 
to their mean-field interaction. Therefore, we have used a 
Thomas-Fermi profile for the condensate to calculate the collision 
integral in the expression for the Keldysh self-energy in 
Eq.~(\ref{sigmak}). As a result the Keldysh self-energy turns out 
not to be constant over the size of the condensate in this limit, 
contrary to our assumption in the derivation of the stochastic 
equations for the variational parameters. To compensate for this 
effect, we have calculated $\hbar \Sigma^{\rm K}$ by taking a 
volume average of $\hbar \Sigma^{\rm K} (\bx)$ over the size of 
the condensate. We report our results  for the number of 
condensate atoms, the frequencies and the damping rates as a 
function of the reduced temperature $T/T_{\rm BEC}$, where 
$T_{\rm BEC}$ is the critical temperature for an ideal Bose gas. 
 
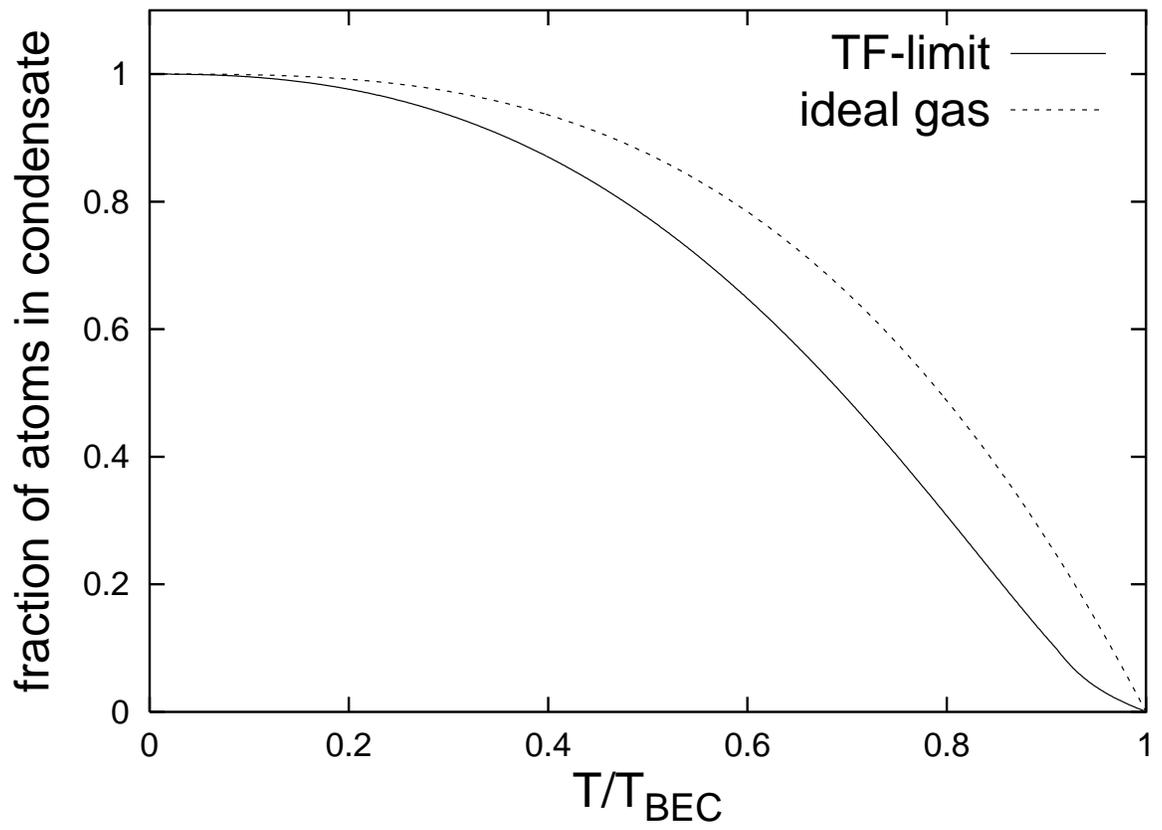
\begin{figure} 
\caption{\narrowtext 
   The solid line gives the condensate fraction for a $^{87}$Rb gas of $2 
   \times 10^6$ atoms in an isotropic trap with $\omega_0/2 \pi=10$ Hz. 
   The dashed line corresponds to the 
   ideal Bose gas in the thermodynamic limit.} 
\label{figurencond} 
\end{figure}

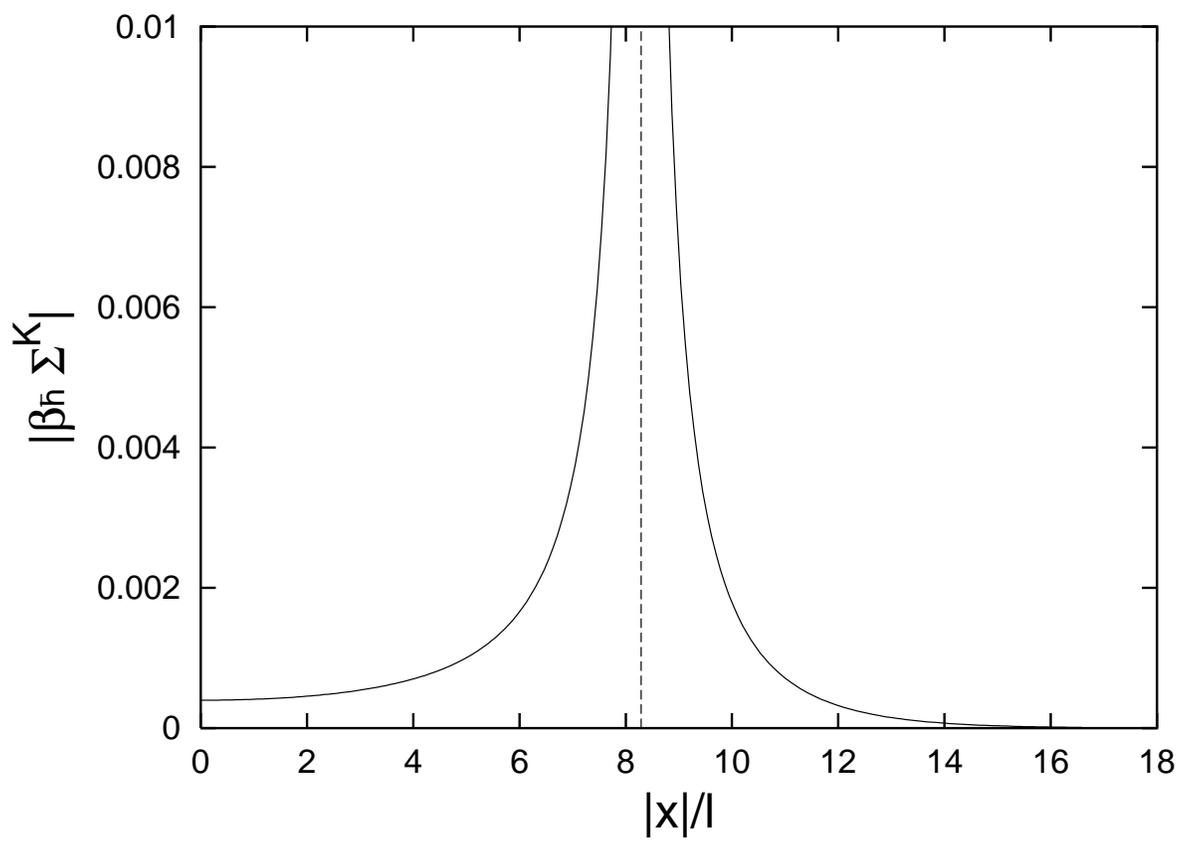
\begin{figure} 
\caption{\narrowtext 
   The dimensionless quantity $| \beta \hbar \Sigma^{\rm K} (\bx)|$. At 
   $|\bx|=R_{\rm TF}$ there is a divergence, when the collision integral in 
   Eq.~(\ref{sigmak}) is calculated in the Thomas-Fermi limit. 
   This is indicated by the dashed line. The 
   calculation is performed for $T=0.5 T_{\rm BEC}$ with the same parameters 
   as in Fig.~\ref{figurencond}} 
\label{figsigmak} 
\end{figure}

The procedure for calculating the number of atoms in the condensate as a 
function of temperature is as follows. For a given number of 
condensate atoms $N_0$ we first calculate the average condensate density profile and the 
chemical potential from the time-independent Gross-Pitaevskii equation. 
In the Thomas-Fermi limit the average condensate 
density profile is given by 
\begin{equation} 
\label{TFprofile} 
  | \langle \phi (\bx) \rangle |^2 = \frac{1}{T^{\rm 2B}} 
      \left( \mu - V^{\rm ext} (\bx) \right), 
\end{equation} 
with a condensate chemical potential 
\begin{equation} 
\label{TFchempot} 
  \mu = \frac{\hbar \omega_0}{2} \left( 15 N_0 a/l \right)^{2/5}, 
\end{equation} 
where $l=\sqrt{\hbar/m \omega_0}$ is the harmonic oscillator 
length. Clearly, the condensate density can not be negative, so 
Eq.~(\ref{TFprofile}) is only valid if the condensate density is 
positive, otherwise it should be taken equal to zero. At $|\bx| = l \sqrt{\mu/ 
\hbar \omega_0} \equiv R_{\rm TF}$ the condensate density 
is equal to zero. Next we calculate the number of atoms in the 
thermal cloud with the value of the chemical potential determined 
by Eq.~(\ref{TFchempot}) using 
\begin{equation} 
  N_{\rm thermal} = \int d \bx \ \int \frac{d {\bf k}}{(2 \pi)^3} 
           N (\epsilon (\bx,{\bf k})), 
\end{equation} 
with $N(\epsilon(\bx,{\bf k}))$ given by Eqs.~(\ref{bosedistr}) 
and (\ref{energythermparttimedep}). We repeat these steps for a 
variable number of condensate atoms until $N_0 + N_{\rm thermal} 
= N_{\rm total}$. The result of the calculation is shown in 
Fig.~\ref{figurencond}, together with the result for an ideal Bose 
gas in the thermodynamic limit. Using this result for $N_0$ as a 
function of the temperature $T$, we have calculated the Keldysh 
self-energy as a function of temperature using the expression in 
Eq.~(\ref{sigmak}) \cite{collintegral}. In Fig.~\ref{figsigmak} 
the function $|\beta \hbar \Sigma^{\rm K} (\bx)|$ is shown, for 
$T=0.5 T_{\rm BEC}$. It is clear from this figure that the Keldysh 
self-energy is not constant over the size of the condensate, and 
even diverges in the Thomas-Fermi approximation at $|\bx|=R_{\rm 
TF}$. The equilibrium values of the variational parameters were 
calculated using Eq.~(\ref{qzeroth}). Subsequently, the various 
parameters were calculated from Eqs.~(\ref{R}), (\ref{alphaj}), 
(\ref{dampingj}) and (\ref{qfreq}). The complex frequency of the 
quadrupole was calculated from Eq.~(\ref{quadfreq}), and the 
complex frequency of the monopole mode was calculated from the 
corresponding analytical expression, which we have omitted here. 
 
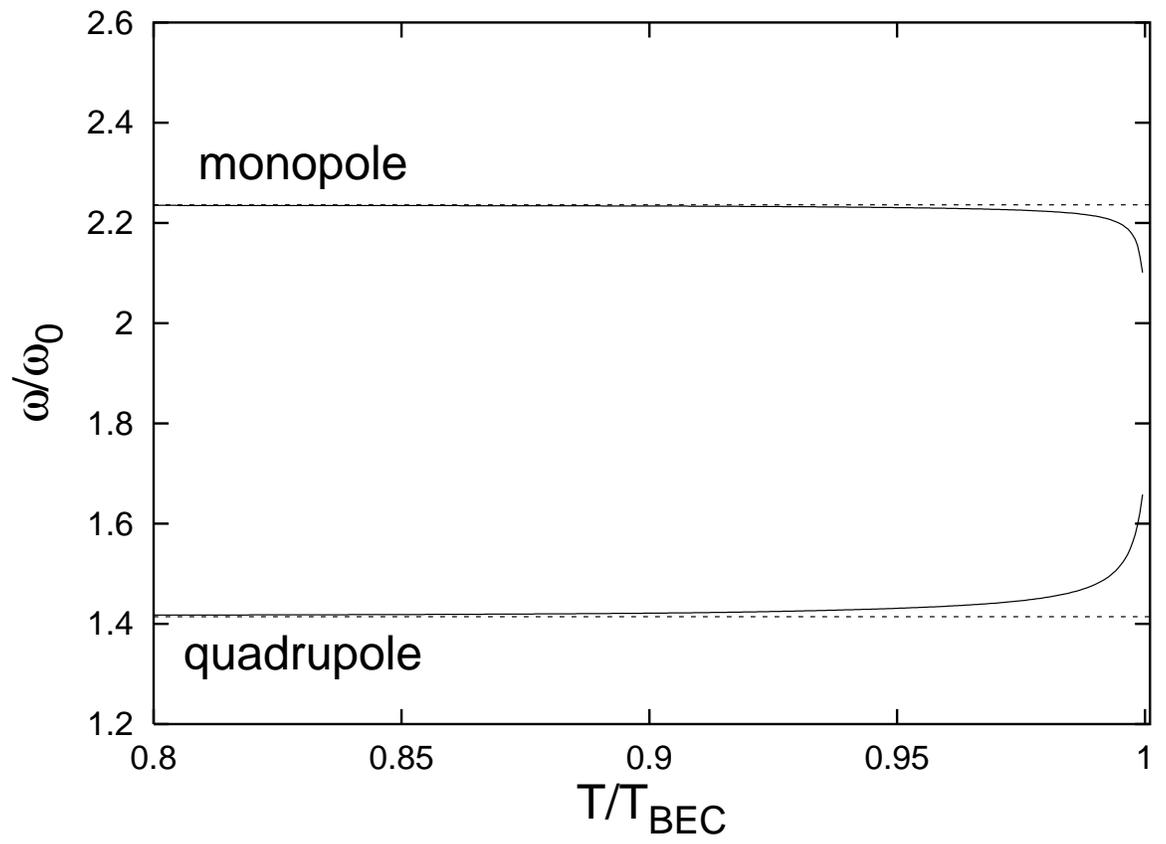
\begin{figure} 
\caption{\narrowtext 
   Frequencies for both the monopole and quadrupole mode as a function of the 
   temperature. The dashed lines indicate the zero-temperature results found 
   by Stringari \protect \cite{stringari}, i.e., 
   $\omega_{\rm quad}= \sqrt{2} \omega_0$ and $\omega_{\rm mono} = \sqrt{5} \omega_0$. 
   The plot starts at $T=0.8 T_{\rm BEC}$ since for smaller temperatures the 
   deviation from these values is neglegible. The parameters are the same 
   as in Fig.~\ref{figurencond}. 
   } 
\label{figisofreq} 
\end{figure}

The results for the frequencies are presented in 
Fig.~\ref{figisofreq}. The dashed lines are the $T=0$ frequencies 
obtained by Stringari \cite{stringari}, in the Thomas-Fermi limit. 
Since the calculations are in the collisionless limit, where $| 
\beta \hbar \Sigma^{\rm K} | \ll 1$, the results are essentially 
$T=0$ results for a variable number of condensate atoms. In the 
Thomas-Fermi limit these frequencies are independent of the 
number of atoms in the condensate, but below $N_0 \simeq 
 10^4$ condensate atoms, where the Thomas-Fermi approximation starts to 
 break down, the frequencies deviate from the $T=0$ results, as 
 is seen in Fig.~\ref{figisofreq}. 
Although for the monopole mode the full expression for the 
frequency and damping involves the parameters $\Gamma$, 
$\alpha_r$ and $\alpha$, which are related to the fluctuations in 
the number of condensate atoms due to the excitation of a 
collective mode, this hardly affects the results for the 
frequencies and damping of this mode. We therefore conclude that 
the fluctuations in the number of condensate atoms during the 
excitation of a collective mode hardly affect the damping and 
frequency of this mode, and that the expression in 
Eq.~(\ref{monofreq}) is valid as long as $|\beta \hbar 
\Sigma^{\rm K} | \ll 1$. 
 
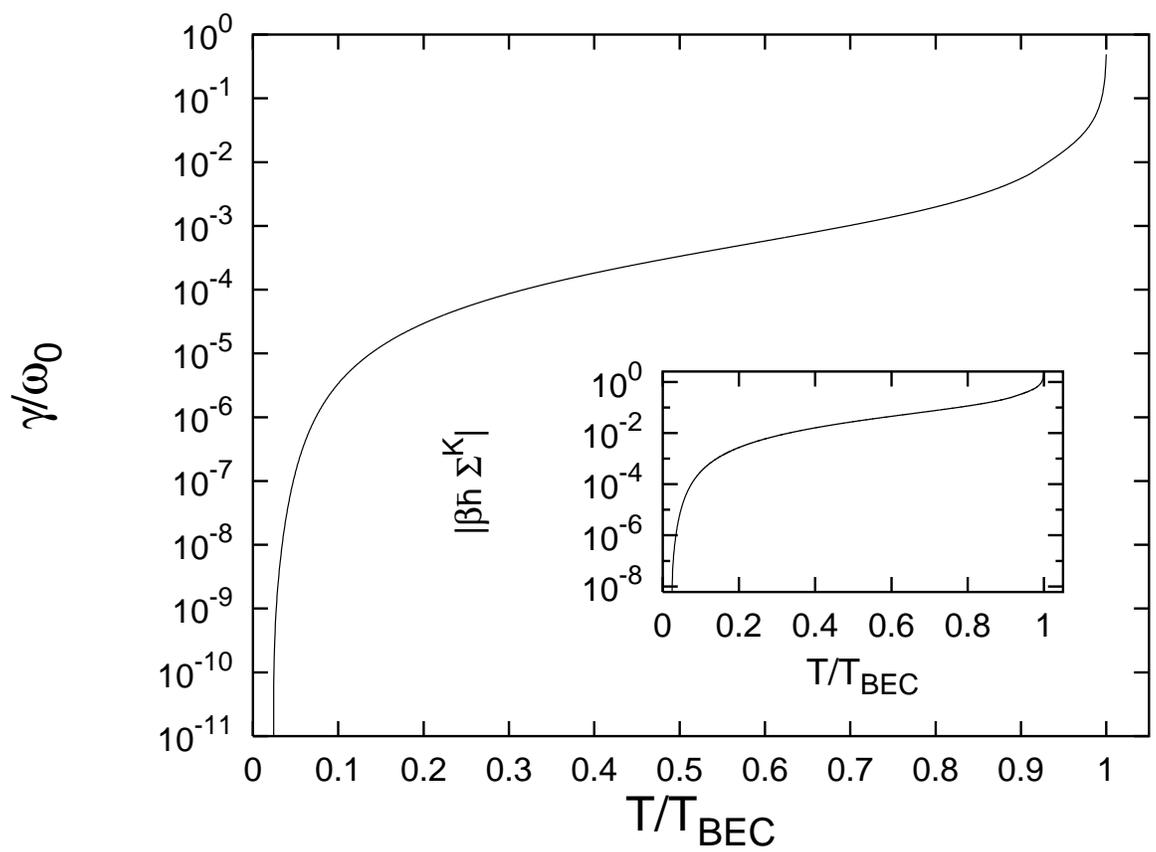
\begin{figure} 
\caption{\narrowtext 
   The damping rate for both the quadrupole and monopole mode for a condensate in an 
   isotropic trap, which is the 
   same for small damping in our variational approximation. The inset shows 
   the dimensionless parameter $|\beta \hbar \Sigma^{\rm K}|$. 
   The parameters are the same 
   as in Fig.~\ref{figurencond}.} 
\label{figisodamp} 
\end{figure}

The results for the damping rate are shown in 
Fig.~\ref{figisodamp}, together with the result for $|\beta \hbar 
\Sigma^{\rm K}|$. Within our variational approximation, the 
damping rates for both the quadrupole and the monopole modes are 
found to be the same in the collisionless regime considered here. 
As clearly seen from Fig.~\ref{figisodamp}, the damping rate 
increases with increasing temperature. This is because the 
density of the thermal cloud becomes larger with increasing 
temperature, and there are therefore more collisions between 
condensate and thermal atoms, which cause the damping. Williams 
and Griffin \cite{williams1} have also calculated the damping of 
the monopole mode for a condensate in a spherical trap in 
presence of a static thermal cloud. These authors have 
generalized the wave equation derived by Stringari 
\cite{stringari} to nonzero temperatures, similar to our result 
in Eq.~(\ref{waveeqn}), albeit that their work does not obey the 
fluctuation-dissipation theorem as mentioned previously. They have 
calculated the damping of the monopole mode in perturbation 
theory, considering the damping as a perturbation parameter. Our 
results for the damping of the monopole mode have the same order 
of magnitude as their results. There are however some qualitative 
differences. We find that the damping rate increases 
 very slowly with temperature for a large temperature regime $T<0.95 T_{\rm BEC}$, 
and then increases dramatically as the temperature approaches the 
critical temperature. Williams and Griffin find that the damping 
rate increases much more gradually with increasing temperature. 
These differences are probably due to the fact that these authors 
take into account that the collision integral in 
Eq.~(\ref{sigmak}) has a position dependence. In Ref. 
\cite{williams2} Williams and Griffin improve upon their 
Thomas-Fermi calculation for the damping rate by using the 
Bogoliubov-deGennes equations that follow from linearizing the 
Gross-Pitaevskii equation with an imaginary term. In this case, 
their results for the damping also show a dramatic increase as the 
temperature reaches the critical temperature and the Thomas-Fermi 
approximation breaks down. In this latter work it is also found 
that the damping for the quadrupole mode and the monopole mode are 
slightly different. The fact that we find that the damping for 
both the monopole mode and the quadrupole mode are equal to first 
order in $\Gamma_r$ is a result of neglect of the spatial 
dependence of the Keldysh self-energy. Williams and Griffin also 
find that there is no first-order correction in the damping to the 
real part of the frequencies, a conclusion consistent with 
Eqs.~(\ref{quadfreq}) and (\ref{monofreq}). 
 
Summarizing, we have calculated the damping and frequencies for 
the quadrupole mode and the monopole mode of a condensate in a 
spherical trap. Our results differ slightly both qualitatively 
and quantitatively from the theoretical results found by Williams 
and Griffin \cite{williams1,williams2}. These differences are 
probably mostly due to the fact that the calculations are 
performed for a large number of atoms in the trap, which implies 
that the collision integral involved in the calculation has a 
significant position dependence, which our variational approach 
does not properly account for. However, for a smaller amount of 
atoms in the trap, we believe that our method should give accurate 
results, and goes in principle beyond the perturbation theory 
considered in Refs. \cite{williams1,williams2}. 
 
\subsubsection{Anisotropic trapping potential} 
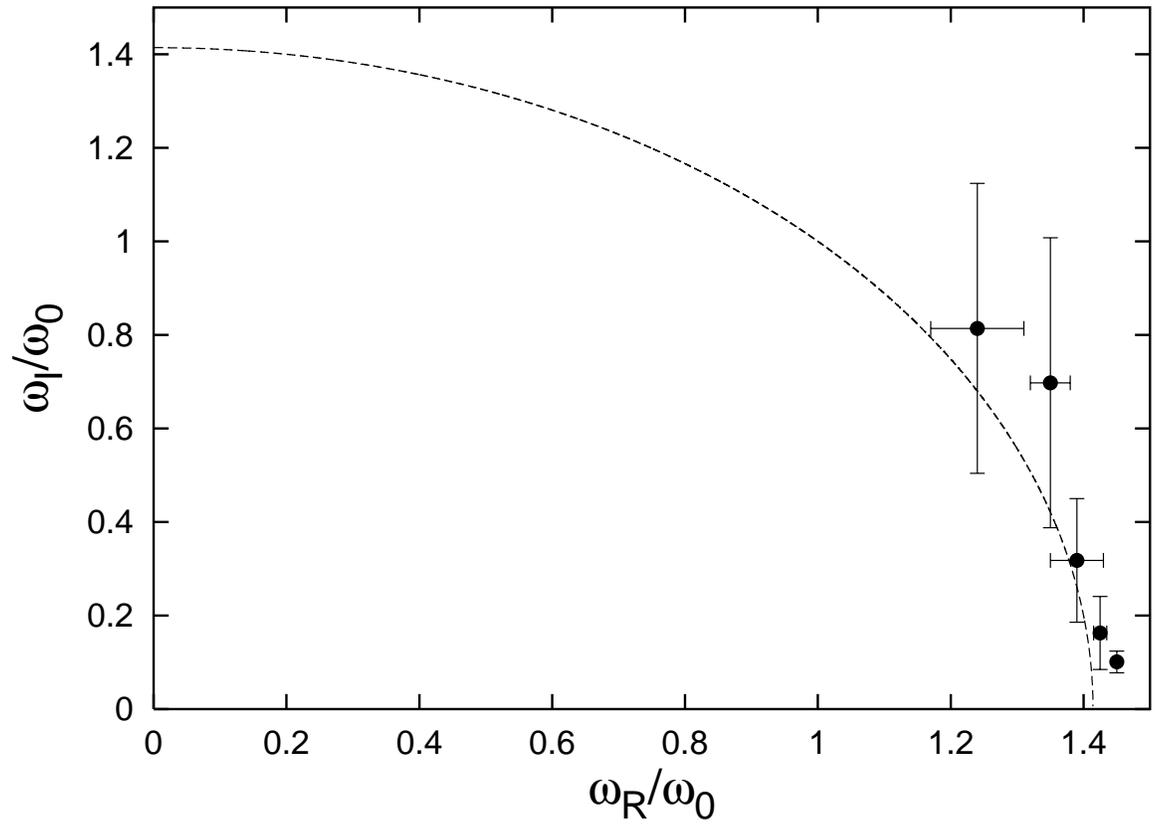
\begin{figure} 
\caption{ \narrowtext 
  The complex $\omega$-plane. In our expression for the 
  frequency and damping rate of the $m=2$ quadrupole mode in 
  Eq.~(\ref{quadfreq}) the complex frequencies lie on a circle of radius 
  $\sqrt{2} \omega_r$. The experimental points taken from 
  Ref.~\protect \cite{jin} are also shown. 
} 
\label{figcircle} 
\end{figure}

We now calculate the frequency of the $m=2$ quadrupole mode, 
where $m$ is the azimuthal quantum number of the angular momentum, 
for the experimental parameters of Jin {\it et al.} \cite{jin}. 
In this experiment, one loads $^{87}$Rb atoms into an anisotropic 
trap, with radial frequency $\omega_r/2 \pi=129$ Hz, and axial 
frequency $\omega_z/2 \pi=365$ Hz. Although the equilibrium shape 
of the condensate is now anisotropic, the expression for the 
frequency of the quadrupole mode found in the isotropic case in 
Eq.~(\ref{quadfreq}), turns out to be also correct for the $m=2$ 
quadrupole mode. The parameters $\Omega_r$, $\Omega_{rr}$, and 
$\Gamma_r$, are now given by 
\begin{eqnarray} 
  \Omega_r \equiv \Omega_x = \Omega_y; \nonumber \\ 
  \Omega_{rr} \equiv \Omega_{xy} = \Omega_{yx}; \nonumber \\ 
  \Gamma_{r} \equiv \Omega_{x} = \Omega_{y}, 
\end{eqnarray} 
which follow from the axial symmetry of the condensate in 
equilibrium. It follows from the expression for the complex 
frequency of the quadrupole mode in Eq.~(\ref{quadfreq}) that the 
complex frequencies lie on a circle of radius $|\omega_{quad}| = 
\sqrt{\Omega_r^2- \Omega_{rr}^2} \simeq \sqrt{2} \omega_r$. To 
test the validity of our expression for the frequency of the 
quadrupole mode, we have plotted the experimental data points 
taken from Ref.~\cite{jin} in the complex $\omega$-plane. In 
Fig.~\ref{figcircle} the result is shown, together with a circle 
of radius $\sqrt{2} \omega_r$. The good quantitative agreement 
with experiment, clearly visible in Fig.~\ref{figcircle}, implies 
that the expression in Eq.~(\ref{quadfreq}) for the frequency and 
damping of the quadrupole mode is correct, even in the 
hydrodynamic regime, where $| \beta \hbar \Sigma^{\rm K} | \gg 
1$. This may in first instance come as a surprise, since our 
variational approximation to the stochastic non-linear 
Schr\"odinger Eq.~(\ref{snlse}) was derived in the collisionless 
regime, where $| \beta \hbar \Sigma^{\rm K} | \ll 1$. Apparently, 
the relation $|\omega_{\rm quad}| \simeq \sqrt{2} \omega_r$ is 
also valid in the hydrodynamic regime. This can be understood 
from the fact that this relation for the complex frequency is 
quite general for a damped harmonic oscillator and we expect on 
general grounds that the quadrupole mode of the condensate can be 
described in this way, both in the collisionless and in the 
hydrodynamic regime. 
 
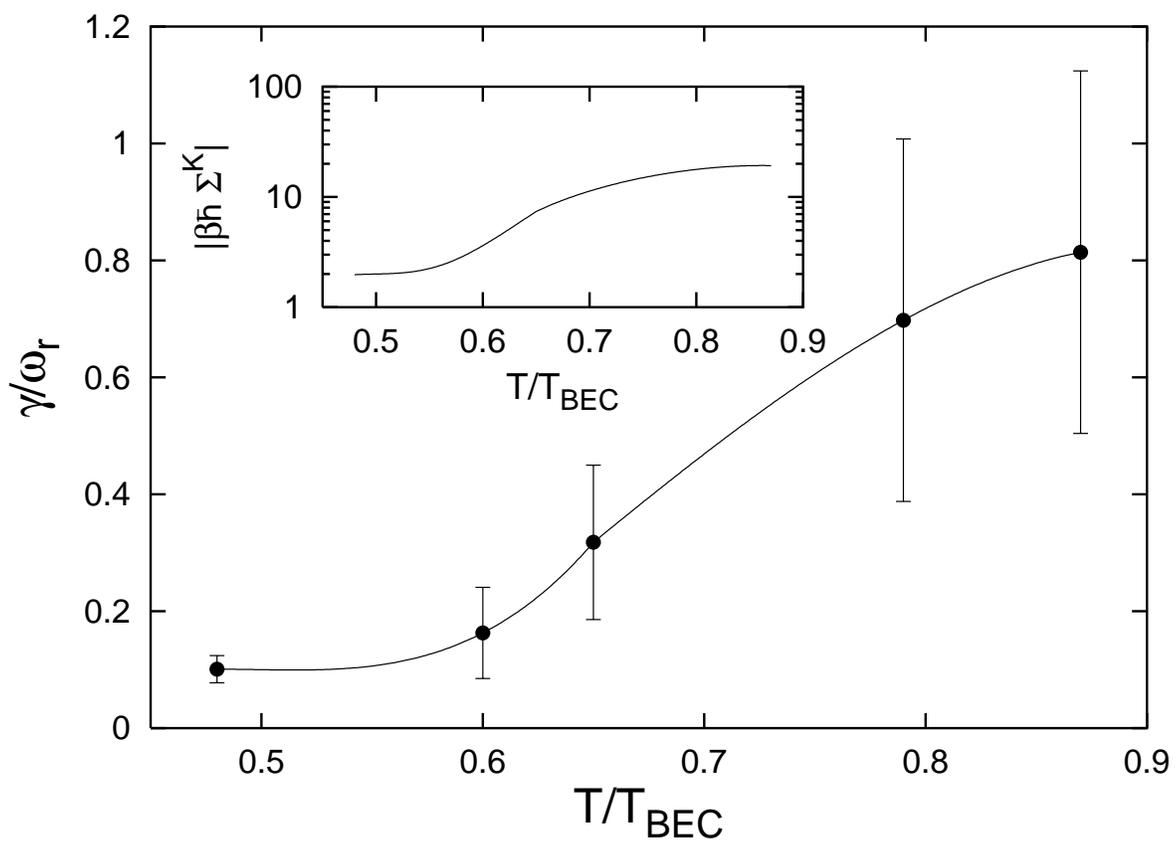
\begin{figure} 
\caption{\narrowtext 
   Fit to the damping rate of the quadrupole mode, as measured by Jin {\it et al.} 
   \protect \cite{jin}. The solid line shows the fit and the experimental points are 
   taken from Ref.~\protect \cite{jin}. The inset 
   shows the value of $|\beta \hbar \Sigma^{\rm K}|$, as 
   calculated from this fit with Eqs.~(\ref{dampingj}), (\ref{qfreq}) 
   and (\ref{quadfreq}). 
} \label{figanisodamp} 
\end{figure}

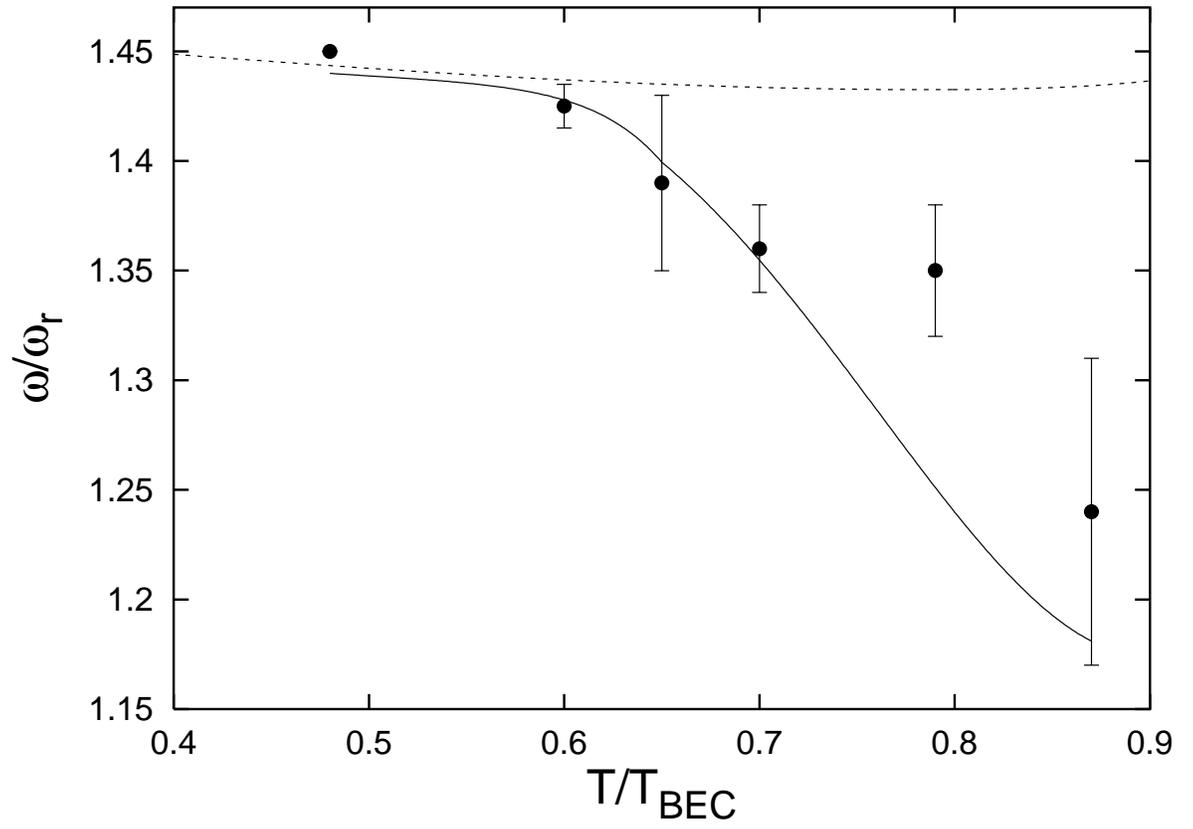
\begin{figure} 
\caption{\narrowtext 
   The frequency of the quadrupole mode as a function of temperature, 
   calculated with Eq.~(\ref{quadfreq}), 
   by using the fit shown in Fig.~\ref{figanisodamp}. 
   The dashed line show the $T=0$ frequency for a 
   variable number of condensate atoms. The experimental points taken from 
   Jin {\it et al.} \protect \cite{jin} are also shown. 
} \label{figanisofreq} 
\end{figure}

To determine $|\beta \hbar \Sigma^{\rm K}|$ as a function of 
temperature, we have fitted the imaginary part of 
Eq.~(\ref{quadfreq}) to the experimental data for the damping 
presented in Ref.~\cite{jin}. From this fit, we have calculated 
the dimensionless parameter $|\beta \hbar \Sigma^{\rm K}|$, using 
Eqs.~(\ref{dampingj}), (\ref{qfreq}), and (\ref{quadfreq}). The 
results of this fit are presented in Fig.~\ref{figanisodamp}. The 
value for $| \beta \hbar \Sigma^{\rm K} |$ found in this manner 
is then used to calculate the real part of Eq.~(\ref{quadfreq}), 
i.e., the frequency of the collective mode. In both the fitting of 
the damping, and the calculation of the frequency, we used the fit 
presented in Ref.~\cite{usama} to determine the experimental 
values for the number of condensate atoms as a function of the 
temperature. The result for the frequency is presented in 
Fig.~\ref{figanisofreq}, together with the frequency calculated 
from Eq.~(\ref{quadfreq}), with $\Gamma_r=0$, i.e., the 
zero-temperature frequency for a variable number of condensate 
atoms. The experimental points are also shown. In 
Fig.~\ref{figanisofreq}, a good quantitative agreement with the 
experimental results is found. Fig.~\ref{figanisofreq} also shows 
clearly that at nonzero temperatures the damping seriously 
affects the frequency of the quadrupole mode, as also found by Al 
Khawaja and Stoof \cite{usama}.  However, a microscopic 
calculation of $\hbar \Sigma^{\rm K}$ in the Thomas-Fermi limit 
for the experimental conditions of interest, by means of a volume 
average of Eq.~(\ref{sigmak}) over the size of the condensate, 
turns out to give values for $|\beta \hbar \Sigma^{\rm K}|$ which 
are approximately one order of magnitude too small to explain the 
experimental data for the quadrupole mode. There are several 
possible reasons for this discrepancy. One possible reason is 
that the calculation of $\hbar \Sigma^{\rm K}$ as an average over 
the size of the condensate of $\hbar \Sigma^{\rm K} (\bx)$ is not 
a good approximation. Since $\hbar \Sigma (\bx)$ becomes large 
compared to $\hbar \Sigma^{\rm K} ({\bf 0})$ at the edges of the 
condensate, where the depletion of the thermal cloud due to the 
condensate's mean field is relatively small, the position 
dependence of $\hbar \Sigma^{\rm K} (\bx)$ is of importance, in 
particular since the density fluctuations occur for the quadrupole 
mode precisely near the edges of the condensate. This is clearly 
shown in Fig.~\ref{figsigmak}, which shows that the Keldysh 
self-energy even diverges at $|\bx|=R_{\rm TF}$ in the 
Thomas-Fermi limit. With respect to this remark, we refer to 
future work concerning the full numerical solution of the average 
of the Langevin equation in Eq.~(\ref{snlse}), to investigate the 
importance of the position dependence of $\hbar \Sigma^{\rm K} 
(\bx)$, and a comparison of this numerically exact approach to 
the variational method developed here. Another possible reason 
for the discrepancy with the experimental results, is the 
presence of other sources of damping, such as Landau and Beliaev 
damping, which have not been included in our calculations.

\subsection{Condensate growth and collapse} 
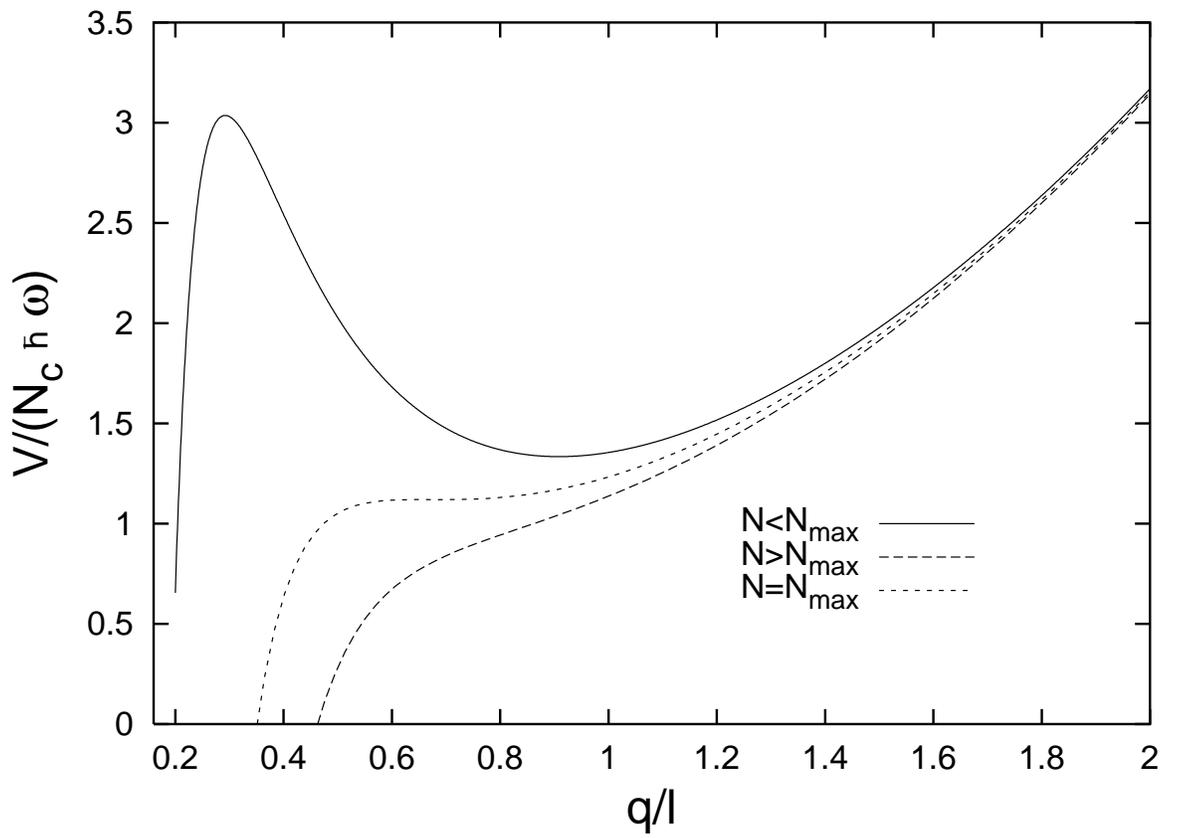
\begin{figure} 
\caption{\narrowtext 
  The condensate energy as a function of the condensate width in the 
  gaussian approximation for three different values of the condensate 
  atom number. If the number of atoms in the condensate is larger than 
  $N_{\rm max}$ the condensate is unstable, otherwise a metastable 
  condensate is possible. 
 } 
\label{figpotq} 
\end{figure}

Although so far we have focused on repulsive interactions, and thus a 
positive scattering length $a$, we consider in this section the case where 
the scattering length is negative. In this case the condensate energy in 
the gaussian approximation becomes 
\begin{equation} 
\label{potentialnegativea} 
   V ({\bf q},N_{\rm c}) = \sum_j \left( \frac{N_{\rm c} \hbar^2}{4 m q_j^2} + \frac{1}{4} m 
  N_{\rm c} \omega_j^2 q_j^2 
                       \right) 
  - \frac{|a| \hbar^2 N_{\rm c}^2}{\sqrt{2 \pi} m q_x q_y q_z}. 
\end{equation} 
From this potential it can easily be seen that there is only a 
metastable condensate possible if the number of atoms in the 
condensate is smaller that a certain critical value. This is 
illustrated in Fig.~\ref{figpotq}, where we show the potential in 
Eq.~(\ref{potentialnegativea}), for several values of the number 
of atoms in an isotropic condensate. For an isotropic trap, the 
maximum condensate number $N_{\rm max}$ turns out to be given by 
the condition 
\begin{equation} 
  N_{\rm max} \frac{|a|}{l} <  \frac{2 \sqrt{2 \pi}}{5^{5/4}} 
  \simeq 0.67. 
\end{equation} 
If the number of atoms in the condensate is above this value, the 
potential in Eq.~(\ref{potentialnegativea}) has no (meta)stable 
minima. If the number of atoms is smaller than $N_{\rm max}$, the 
potential has a metastable minimum, and the condensate can start 
to collapse by overcoming the metastable energy barrier by either 
macroscopic quantum tunneling or thermal fluctuations 
\cite{henk3}. The stability condition for the condensate, found 
by a full numerical solution of the Gross-Pitaevskii, turns out to 
be $N_{\rm c} a/l < 0.58$ \cite{ruprecht}, and thus we see that 
the gaussian approximation is only 16 \% off. 
 
The first experiments on Bose-Einstein condensation in a gas with 
attractive interactions where performed with $^7$Li \cite{curtis}, 
which has a negative scattering length of $a \simeq -1.45$ nm. In 
these experiments, the gas is evaporatively cooled below the 
critical temperature, which causes the condensate to undergo 
several growth and collapse cycles before relaxing to a metastable 
equilibrium \cite{cass1}. Because of the stochastic initiation of 
the collapse, one could not make a sequence of destructive 
measurements. However, a statistical analysis revealed that 
during a collapse the number of condensate atoms is not reduced 
to zero, but that the collapse stops, presumably because of 
elastic and inelastic collisional loss processes, when the number 
of atoms in the condensate is about 200 \cite{cass2}. In a recent 
experiment, one was able to make a sequence of destructive 
measurements by (partially) dumping the condensate by a 
two-photon pulse \cite{randy}, and thus start each measurement 
with approximately the same initial number of condensate atoms. 
 
Using the Langevin equations for $q_j(t)$, and the stochastic rate 
equation for the number of atoms in the condensate, 
we are able to describe this experiment. To do so in the easiest way, we want to model the thermal 
cloud by an equilibrium Bose distribution of a noninteracting gas. 
However, numerical solutions of the quantum 
Boltzmann equation for these experiments have shown that the thermal cloud is not equilibrium, 
but can be well modeled by a distribution function given by \cite{cass2} 
\begin{equation} 
\label{noneqdistr} 
  f (\epsilon) = 
    \frac{\exp[\beta (\mu'-\mu)]}{\exp[\beta(\epsilon - \mu)]-1} 
    \equiv 
    \frac{A}{\exp[\beta(\epsilon - \mu)]-1}. 
\end{equation} 
At high energies, $f$ has the form of a Boltzmann distribution 
with chemical potential $\mu'$ and temperature $1/k_{\rm 
B}\beta$. At low energies $f$ has the form 
$f(\epsilon)=A/\beta(\epsilon-\mu)$, which is precisely the 
low-energy tail of a Bose-distribution with effective temperature 
$A/k_{\rm B}\beta$. Therefore, we conclude that 
Eq.~(\ref{fdtheorem}) is too a good approximation still valid for 
the distribution function given by Eq.~(\ref{noneqdistr}), if we 
make the replacement $\beta \to \beta/A$ in the operator on the 
right hand side of Eq.~(\ref{fdtheorem}).  In this manner we obey 
the fluctuation-dissipation theorem, and have also accounted for 
the fact that the distribution function of the thermal cloud is a 
nonequilibrium distribution function. We expect that this 
approximation will give quantitatively correct results for the 
condensate growth rate, since it is particularly good for the 
low-lying energy levels, which dominate the condensate growth. 
 
Before comparing to the experimental data reported in 
Ref.~\cite{randy}, we discuss some aspects of the numerical 
solutions of the stochastic rate equation coupled to the Langevin 
equation, for a condensate which has initially no atoms. In the 
experiment performed by Gerton {\it et al.} this would correspond 
to the situation where the condensate is dumped completely. The 
stochastic rate equation in Eq.~(\ref{rateeqn}) is well suited 
for this purpose, since it also contains fluctuations, which 
initiate the growth in this case. Without these fluctuations, the 
growth rate of the condensate would never become nonzero. These 
initial conditions for condensate growth are different from the 
experiments conducted by Miesner {\it et al.} \cite{miesner}, in 
which the condensate growth is observed after evaporatively 
cooling the gas. In this case, the ground state already has a 
rather large nonzero occupation number above the critical 
temperature, which causes a growth process dominated by 
Bose-stimulation. Therefore, for a theoretical description of the 
condensate growth it is not so essential to include fluctuations 
in this case \cite{michiel2,crispin}. We perform our simulations 
for the experimental conditions reported in Ref.\cite{randy}. The 
trap frequencies are given by $\omega_r/2 \pi=151$ Hz and 
$\omega_z/2 \pi=131.5$ Hz. We consider a thermal cloud with 
$70000$ atoms at a temperature $T=170$ nK. The parameter $A$ is 
taken equal to $4$. These values correspond to typical 
experimental conditions \cite{ionut}. The Keldysh self-energy 
$\hbar \Sigma^{\rm K}$ is calculated using Eq.~(\ref{sigmak}), 
with the nonequilibrium distribution function 
 $f (\epsilon)$ given by Eq.~(\ref{noneqdistr}). The mean-field effects 
of the condensate on the thermal cloud are neglected, an 
approximation which will certainly be valid in the initial stage 
of the condensate growth, when the condensate is small. Moreover, 
we take the chemical potential of the condensate $\langle 
\mu_{\rm c} \args \rangle$ in Eq.~(\ref{sigmak}) equal to zero. 
Since the density of thermal atom will be the largest in the 
center of the trap, and the condensate is small in this case, we 
do not perform an average to calculate $\hbar \Sigma^{\rm K}$, 
but simply take $\hbar \Sigma^{\rm K} = \hbar \Sigma^{\rm K} 
({\bf 0 })$. 
 
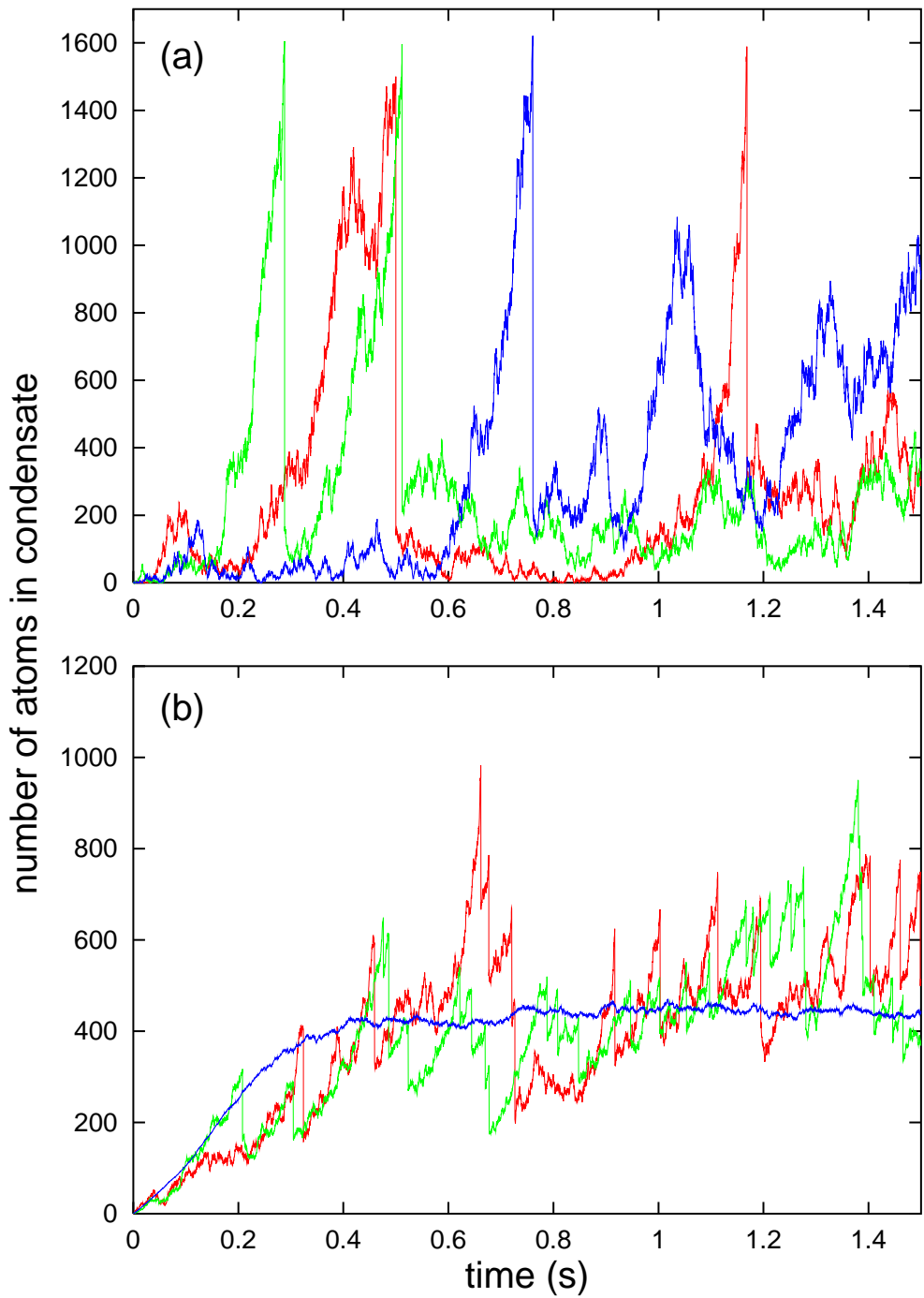
\begin{figure} 
\caption{\narrowtext 
  (a) Growth-collapse curves of a $^7$Li condensate, and (b) their averages. 
  The colored lines in (a) display 
  the number of condensate atoms for solutions of the Langevin equation for 
  $q_j(t)$, 
  coupled to the stochastic rate equation for $N_{\rm c}(t)$, 
  for three different realizations of the noise. 
  In (b) the red line correspond to an average over $5$ realizations, 
  the green line to $10$, and the blue line to an average over $1000$ 
  different realizations of the noise. 
  The simulations are done for a thermal cloud of 
  $70000$ atoms with $T=170$ nK, and $A=4$. 
 } 
\label{figsolutions} 
\end{figure}

We solve the Langevin equations coupled to the stochastic rate 
equation, using standard numerical techniques for stochastic 
differential equations \cite{drummond}. Since the initial number 
of condensate atoms is equal to zero, we put at time $t=0$ the 
values of the gaussian variational parameters equal to 
$q_j=\sqrt{\hbar/m \omega_j}$, which is their equilibrium value 
in the limit where the number of condensate atoms approaches 
zero. A slight subtlety arises in the use of the stochastic rate 
equation. One has to realize that the chemical potential of the 
thermal cloud in this equation is measured with respect to the 
energy of the lowest excited level, since we want the 
distribution function to describe the non-condensed atoms only. 
This means that we should use in the rate equation the chemical 
potential found in matching the distribution function in 
Eq.~(\ref{noneqdistr}) to the number of thermal atoms, and add the 
energy of the lowest excited level to it. This is immediately 
clear when we write the number of atoms in the thermal cloud as a 
sum over occupation numbers, instead of an integral over energy. 
Fig.~\ref{figsolutions} shows the results of our simulations. In 
Fig.~\ref{figsolutions}~(a), we plot the number of condensate 
atoms as a function of time, for the solutions of the Langevin 
equations and the stochastic rate equation for three different 
realizations of the noise. We assume that during the growth and 
subsequent collapse of the condensate, 
 the distribution function of the thermal cloud is not 
affected. The maximum number of atoms in the condensate is for the 
parameters under consideration here equal to $N_{\rm max} \simeq 
1470$ atoms, within the gaussian approximation. This means that 
during one growth-collapse cycle the number of atoms in the 
thermal cloud is reduced by only approximately $2$ \%, and 
therefore this approximation seems valid for the description of 
one growth-collapse curve. Once a collapse is initiated by the 
noise in the Eqs.~(\ref{qlangevin}) and (\ref{rateeqn}), we model 
the collapse by putting the number of condensate atoms 
instantaneously equal to a gaussian random number with a mean 
value of $200$ and a deviation of $40$, which corresponds to the 
$20$ \% systematic uncertainty reported in \cite{cass2}, and the 
variational parameters $q_j (t)$ equal to their corresponding 
equilibrium values, given by Eq.~(\ref{qzeroth}). The collapse 
occurs on a time-scale ${\mathcal O} (1/\omega)$, which is much 
faster than the time-scale on which the condensate grows. Since 
we are interested in the growth process here, and not in the loss 
process which stops the collapse, this appears a reasonable way to 
model the collapse. Fig.~\ref{figsolutions} (a) clearly shows 
that when the number of condensate atoms approaches the maximum 
number $N_{\rm max}$, the condensate tends to collapse. We found 
that the collapse is initiated stochastically by the fluctuations 
in the number of atoms in the condensate, which cause density 
fluctuations that cause the condensate to overcome the 
macroscopic energy barrier and start the collapse. Since our 
description only includes thermal fluctuations, we may ask if 
macroscopic quantum tunneling might be of importance. However, 
previous work has shown that decay by thermal fluctuations is the 
main decay mechanism for the temperatures of interest 
\cite{cass1}. In Fig.~\ref{figsolutions} (b) we plot the number of 
atoms in the condensate as a function of time, averaged over 
different realizations of the noise.  The red line is an average 
over $5$ different realizations of the noise. A growth-collapse 
signature is still visible in this curve, although the stochastic 
growth process and initiation of the collapse has led to a 
dephasing of the moment of collapse. The green line shows an 
average over $10$ realizations of the noise. Although the initial 
growth is clearly visible in this curve, the collapse can hardly 
be seen from this average, since the noise has led to an almost 
complete dephasing, and the collapse is `averaged out'. Finally, 
the blue curve shows an average over $1000$ realizations of the 
noise. No signature of the collapse is visible in this curve, 
because the averaging leads to a complete dephasing of the moment 
of the collapse. 
 
We now discuss the simulation of the experiments performed by 
Gerton {\it et al.} \cite{randy}. To make a comparison with 
experiment, one has to realize that each data point is obtained 
as an average over $5$ or $10$ individual runs. Since the 
condensate number is probed by means of a destructive 
measurement, each experimental curve should not be viewed as an 
average of curves. Instead, each point is an average over 
different experimental runs, and the time-correlation between 
different experimental points is only caused by the initial 
conditions, which are approximately the same for each 
experimental run. To simulate this experiment by means of a 
numerical solution of the Langevin equations in 
Eqs.~(\ref{qlangevin}) and (\ref{rateeqn}), we therefore have to 
let the numerical solutions evolve up to a certain point in time, 
and then make a numerical measurement. We then average over $5$ 
or $10$ measurements to obtain a data-point and its uncertainty, 
and repeat this procedure at a different measurement time. In 
this way, we are certain that each individual solution of our 
stochastic equations, is done for a different realization of the 
noise. Note that this procedure is very reminiscent of the method 
of Monte Carlo simulation. 
 
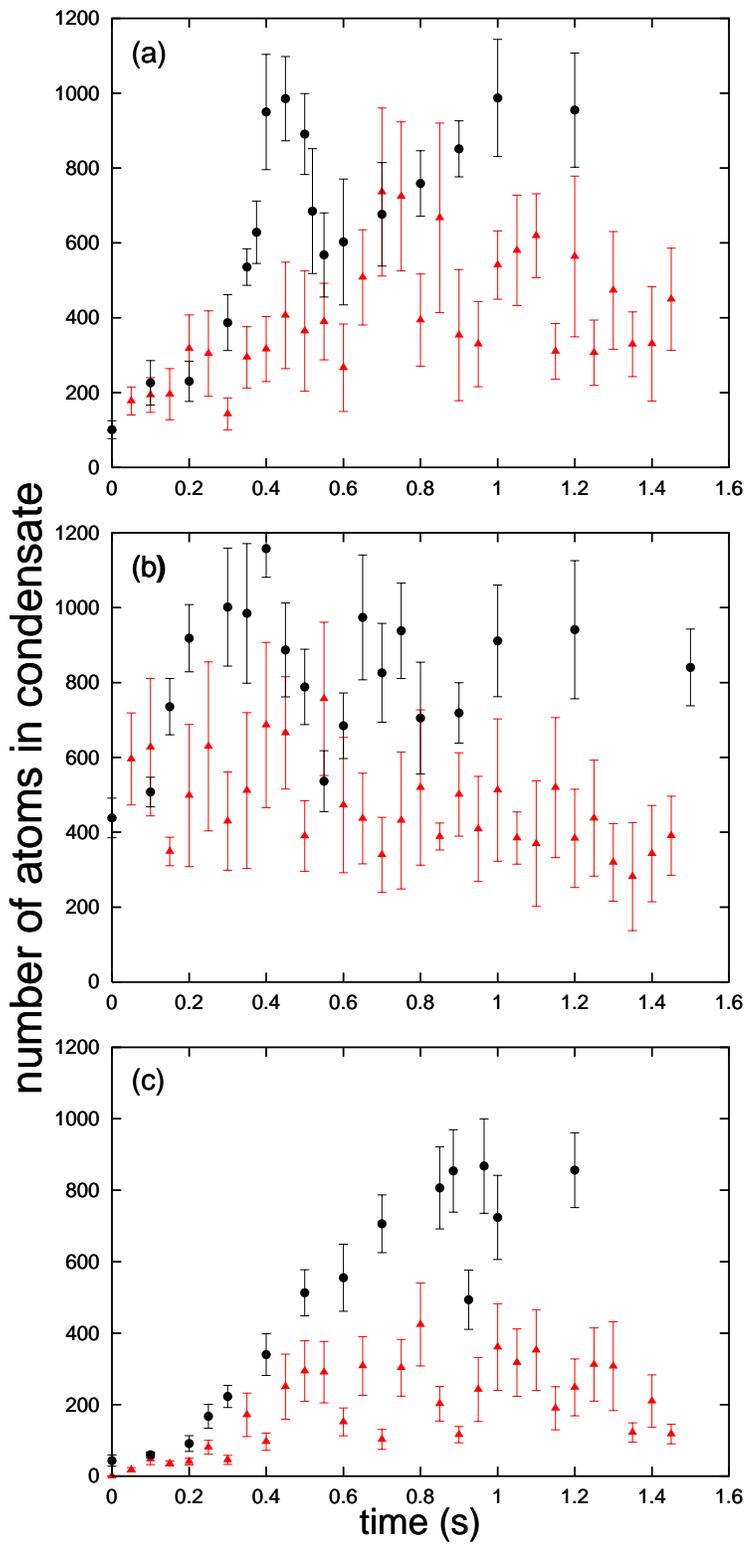
\begin{figure} 
\caption{\narrowtext 
  Simulations of the experiment performed by Gerton {\it et al.} \protect 
  \cite{randy}. The results of the simulations are denoted by red triangles, 
  the experimental data is shown as black circles. In (a) and (b) each 
  data-point of the simulations is an average over $5$ runs, as in the 
  experiment. For (c) $10$ runs per point were done. The error bars in both 
  the experimental data, and the data obtained by the simulations, denote 
  the uncertainty in the mean. 
} \label{figexperiments} 
\end{figure}

We have done simulations for the three different experimental 
situations presented in Ref.~\cite{randy}. The results of our 
simulations are presented in Fig.~\ref{figexperiments}, as red 
triangles. The experimental data-points are also shown, and 
denoted by black circles. The Keldysh self-energy was calculated 
as in the simulations described above, using the averages of the 
full experimental data on the number of atoms in the thermal cloud 
and their temperature \cite{ionut}. For the parameter $A$ of the 
nonequilibrium distribution function in Eq.~(\ref{noneqdistr}) we 
use the average of the fits obtained by Gerton {\it et al.} 
\cite{ionut}. For Fig.~\ref{figexperiments} (a) and (c) this 
corresponds to a thermal cloud of approximately $65000$ atoms at 
a temperature of $T = 170$ nK. The parameter of the 
nonequilibrium distribution is equal to $A = 4$ for these 
simulations. For Fig.~\ref{figexperiments}~(b) the thermal cloud 
contains approximately $ 100000$ atoms at a temperature of $T = 
200$ nK. The parameter $A = 2$ in this case. For 
Fig.~\ref{figexperiments}~(a) and (b) the averages are taken over 
$5$ runs for each data-point, whereas for (c) $10$ runs are used. 
The error bars denote the uncertainty in the average. In 
Fig.~\ref{figexperiments}~(a) and (b) the initial number of atoms 
was take equal to $N_{\rm c}(0)=100$, and $N_{\rm c}(0)=438$ 
respectively. For Fig.~\ref{figexperiments}~(c) we have taken 
$N_{\rm c}(0)=0$, since the condensate was in this case dumped 
completely to within the experimental resolution \cite{randy}. 
 
The results of our simulations presented in 
Fig.~\ref{figexperiments} shows good agreement with experiment for 
the initial stage of the growth, where the condensate is small. 
In particular, Fig.~\ref{figexperiments}~(a) show good agreement 
in the initial stage where $N_{\rm c} < 400$ atoms, whereas 
Fig.~\ref{figexperiments}~(c) shows good agreement in the regime 
where $N_{\rm c} < 600$ atoms. This is to be expected, since the 
gaussian {\it ansatz} is a very good approximation for a small 
number of atoms in the condensate, whereas it becomes worse for a 
larger number of atoms in the condensate. The fact that the error 
bars of the experimental data points have the same order of 
magnitude as the error bars on our simulations, indicates that 
the fluctuations, i.e., the noise in our stochastic equation, have 
indeed the correct order of magnitude. As mentioned in the 
discussion of the individual solutions of our stochastic 
equations, we find that the collapse is initiated by fluctuations 
in the number of atoms, and that these fluctuations thus lead to 
a dephasing of the moment of the collapse. In principle also the 
fluctuations in the initial number of atoms in the condensate 
lead to dephasing. However, the uncertainty in $N_{\rm c} (0)$ is 
small compared to the uncertainty in $N_{\rm c} (t)$ at later 
times, and we therefore conclude that fluctuations in the initial 
conditions for the condensate are presumably less important for 
an understanding of the dephasing of the moment of the collapse. 
Moreover, there are also fluctuations in the properties of the 
thermal cloud for each individual experimental run, which we have 
not taken into account. With respect to this point we also note 
that our method does not display the saturation in the growth 
rate, observed in numerical solutions of the quantum Boltzmann 
equation \cite{randy}. This effect is also observable in the 
experimental data in Fig.~\ref{figexperiments}~(c), where the 
growth is exponential in the first stage, but then turns linear. 
This saturation in the growth rate is caused by the fact that the 
condensate mostly grows from the low-lying excited states, which 
in turn have to be feeded by collisions in higher energy states, 
which are not Bose enhanced. Since in our simulations the thermal 
cloud is taken to be static, our simulations do not display this 
saturation effect. In conclusion, we like to point out that, to 
make a sensible quantitative comparison to the experimental 
results in the whole time-domain, we have to compare the 
converged averages of both the experimental runs and the 
theoretical simulations. This is because of the fact that the 
fluctuations are so large, that each individual growth curve can 
differ substantially, as seen from Fig.~\ref{figsolutions}~(a). 
In turn, this leads to averages that can differ qualitatively, 
depending on the number of runs one averages over, as also 
clearly seen in Fig.~\ref{figsolutions}~(b). 
 
\section{Conclusions} 
We have presented  a Fokker-Planck equation that describes the 
nonequilibrium dynamics of an atomic Bose-Einstein condensate. We 
have discussed an approximation to this Fokker-Planck equation, 
which assumes that the thermal cloud is close to equilibrium. Its 
corresponding Langevin equation has the form of a stochastic 
nonlinear Schr\"odinger equation with complex gaussian noise. 
Both the Fokker-Planck equation, and the Langevin equation obey 
the fluctuation-dissipation theorem, which ensures that the 
condensate relaxes to the correct equilibrium. We have also 
presented the hydrodynamic formulation corresponding to this 
stochastic non-linear Schr\"odinger equation, in which the 
condensate is described in terms of its density and its phase. 
These turn out to obey a stochastic continuity equation and a 
stochastic Josephson equation, respectively. To make analytical 
progress, we have then extended the variational calculus, 
commonly applied to the Gross-Pitaevskii equation, also to the 
case of the stochastic non-linear Schr\"odinger equation. The 
equations of motion for the variational parameters turn out to be 
identical to the Langevin equations describing the Brownian motion 
of a particle in a potential. These equations are then coupled to 
a stochastic rate equation for the number of atoms in the 
condensate. We have applied these equations to calculate the 
damping and frequencies of the collective modes of the 
condensate, and to obtain a description of the growth-collapse 
curve of a condensate with attractive interactions. However, 
there are much more applications possible with the variational 
method presented here. With a slight extension, it can also be 
used to calculate the frequency and damping of the scissor modes 
of the condensate \cite{foot,eugene}, at nonzero temperatures. 
 Moreover, applying the method to a Thomas-Fermi density 
profile, we can obtain a simple description of the growth of a 
condensate with repulsive interactions. The treatment of the 
dissipative dynamics of vortices and other topological 
excitations such as skyrmions \cite{usama2}, is also feasible 
within this variational method. 
 
The only quantity that characterizes the thermal cloud is, in our 
approach, the Keldysh self-energy $\hbar \Sigma^{\rm K} \args$. 
The parameter that enters the equations of motions for the 
variational parameters, turns out to be some spatial average of 
this quantity. In our calculations presented here, we have taken 
an average over the size of the condensate, when calculating the 
frequency and damping of the collective modes. Future work will 
include a numerical solution of the stochastic non-linear 
Schr\"odinger equation, to investigate the importance of the 
spatial dependence of the Keldysh self-energy, which we have not 
taken into account here. Nevertheless, we believe that the 
variational method presented here, provides a satisfying picture 
of the nonequilibrium dynamics of a Bose-Einstein condensate at 
nonzero temperatures. Moreover, as we have shown, calculations 
done within this variational approximation, lead already to a good 
agreement with experiments on collective modes and on condensate 
growth.

\section*{Acknowledgements} 
It is a great pleasure to thank Ionut Prodan and Randy Hulet for 
kindly providing us with their raw experimental data. We also 
thank Usama Al Khawaja for useful remarks, and for providing us 
with some numerical results.

\appendix 
\section{Amplitude and phase variables} The condensate is often 
described in terms of density and phase variables, by making a 
canonical transformation $\phi = \sqrt{\rho} e^{i \theta}$. In 
this appendix we discuss the derivation of the Langevin equations 
of motion for $\rho \args$ and $\theta \args$. For simplicity, we 
first discuss the single-mode version of the probability 
distribution in Eq.~(\ref{probdistr}) in the non-interacting 
case. So we consider a probability distribution for a single-mode 
complex order parameter, which reads 
\begin{equation} 
\label{singlemodeprobdistr} 
  P[\phi^*,\phi;t] = 
    \int^{\phi^*(t)=\phi^*}_{\phi(t)=\phi} d[\phi^*] d[\phi] 
      \exp 
        \left\{ 
      \frac{i}{\hbar} 
        \left( 
          \int_{t_0}^{t} dt' 
            \frac{2}{\hbar \Sigma^{\rm K}} 
        \left| 
          \left( 
            i \hbar \frac{\partial}{\partial t'} 
            + \mu - \mu_{\rm c} + i R 
          \right) \phi (t') 
        \right|^2 
        \right) 
        \right\}. 
\end{equation} 
Physically, this probability distribution describes the 
nonequilibrium dynamics of a non-interacting Bose-Einstein 
condensate, in contact with a thermal cloud characterized by a 
Keldysh self-energy $\hbar \Sigma^{\rm K}$, with inverse 
temperature $\beta$ and a chemical potential $\mu$. The energy 
per particle in the single-mode system is equal to $\mu_{\rm c}$. 
The dissipation $R$ is again related to the Keldysh self-energy 
by the fluctuation-dissipation theorem in Eq.~(\ref{fdtheorem}), 
which in this simple case reads 
\begin{equation} 
\label{fdtheoremsinglemode} 
  i R = - \frac{\beta}{4} \hbar \Sigma^{\rm K} [\mu_{\rm c} -\mu]. 
\end{equation} 
From the previous sections we know that the probability distribution 
$P[\phi^*,\phi;t]$ is generated by the Langevin equation 
\begin{equation} 
\label{singlemodelangevin} 
  i \hbar \frac{\partial \phi (t)}{\partial t} = 
    \left( \mu_{\rm c} - \mu -iR 
    \right) \phi(t) + 
    \eta (t), 
\end{equation} 
where the complex noise has a time correlation function given by 
\begin{equation} 
  \langle \eta^* (t') \eta (t) \rangle = \frac{i \hbar^2}{2} 
    \Sigma^{\rm K} \delta (t'-t). 
\end{equation} 
As explained in the first section, the noise $\eta (t)$ has to be 
interpreted as an Ito process. 
We can again derive the Fokker-Planck equation for $P[\phi^*,\phi;t]$ 
by noting that it is in fact the Schr\"odinger equation in the position 
representation. We will do this in some detail once more, 
to make clear the different steps of the derivation. 
The first step is to determine the momenta conjugate to 
the coordinates $\phi$ and $\phi^*$. Since we have a lagrangian equal to 
\begin{equation} 
  L[\phi^*,\phi]=\frac{2}{\hbar \Sigma^{\rm K}} 
        \left| 
          \left( 
            i \hbar \frac{\partial}{\partial t'} 
            + \mu - \mu_{\rm c} + i R 
          \right) \phi (t') 
        \right|^2, 
\end{equation} 
we can define the momentum conjugate to $\phi$ in the usual way 
\begin{equation} 
  p_{\phi} = \frac{\partial L}{\partial \phi}=\frac{2i}{\hbar \Sigma^{\rm K}} 
  \left( -i \hbar \frac{\partial}{\partial t} - \mu_{\rm c} -i R + \mu \right), 
\end{equation} 
with the complex conjugate expression for $p_{\phi^*}$. The 
second step is to derive the hamiltonian. Although it has in 
principle ordering problems, we overcome these by noting that in 
the path-integral formulation of quantum mechanics we are always 
dealing with a normal ordered hamiltonian. The normal ordered 
hamiltonian, i.e., with the momentum operators positioned left to 
the coordinate operators, is now given by 
\begin{equation} 
  H[p_{\phi},\phi;p_{\phi^*},\phi^*] = p_{\phi} \dot \phi + p_{\phi^*} \dot 
  \phi^* - L [\phi^*,\phi]. 
\end{equation} 
The last step towards the Fokker-Planck equation is to quantize the 
hamiltonian, and to write down the 
Schr\"odinger equation in the position representation. So we have $p_{\phi} 
= - i \hbar \partial/ \partial \phi$, and similarly $p_{\phi^*} 
= - i \hbar \partial/ \partial \phi^*$. The Fokker-Planck equation becomes 
\begin{eqnarray} 
\label{fpsinglemodephi} 
  i \hbar \frac{\partial}{\partial t} P[\phi^*,\phi;t] 
     &=& -\frac{\partial}{\partial \phi} 
        (\mu_{\rm c}-\mu-iR) \phi P[\phi^*,\phi;t] \nonumber \\ 
     &~&~ \frac{\partial}{\partial \phi^*} 
        (\mu_{\rm c}-\mu-iR) \phi^* 
          P[\phi^*,\phi;t] \nonumber \\ 
     && - \half \frac{\partial^2}{\partial \phi^* \partial \phi} 
     \hbar \Sigma^{\rm K} P[\phi^*,\phi;t]. 
\end{eqnarray} 
With the use of the fluctuation-dissipation theorem in 
Eq.~(\ref{fdtheoremsinglemode}) we then show that it has as a 
stationary solution 
\begin{equation} 
\label{singlemodeequilibrium} 
  P[|\phi|, t \to \infty] \propto \exp \left\{-\beta \phi^* (\mu_{\rm c} -\mu) 
  \phi \right\}, 
\end{equation} 
which only depends on the amplitude of $\phi$ and $\phi^*$. 
 
We now repeat the above discussion in terms of amplitude and phase 
variables, defined by $\sqrt{N} e^{i \theta}$. Let us first 
discuss the equilibrium properties we expect in terms of the 
number of particles $N$ and the phase $\theta$. Since the 
transformation to $N$ and $\theta$ has a jacobian equal to zero, 
we can just substitute it into Eq.~(\ref{singlemodeequilibrium}), 
to obtain the equilibrium distribution in terms of the number of 
particles. It is given by 
\begin{equation} 
\label{equiln} 
  P[N; t \to \infty] \propto \exp \left\{-\beta  (\mu_{\rm c} -\mu) N \right\}. 
\end{equation} 
We can easily check that the average number of particles in the single 
mode is in equilibrium given 
\begin{equation} 
  \langle N \rangle \equiv \frac{\int_0^{\infty} dN \ N P[N; t \to 
  \infty]}{\int_0^{\infty} dN \ P[N; t \to \infty]} = [\beta (\mu_{\rm c} - 
  \mu)]^{-1}. 
\end{equation} 
This is precisely the Bose distribution for $\beta (\mu_{\rm c} - 
\mu) \ll 1$. If we do not apply the `classical' approximation to 
the fluctuation-dissipation theorem, as in 
Eq.~(\ref{fdtheoremsinglemode}), but use the exact relation 
\begin{equation} 
  i R = -\frac{1}{2} \hbar \Sigma^{\rm K} 
    \left( 
      1+2 N (\mu_{\rm c}) 
    \right)^{-1}, 
\end{equation} 
instead, we find the Bose distribution as the equilibrium number 
of particles, as expected. For the description of a single-mode 
Bose-Einstein condensate Eq.~(\ref{fdtheoremsinglemode}) is in 
general a good approximation, since $\mu$ is very close to 
$\mu_{\rm c}$ below the critical temperature. Let us now try to 
derive the Fokker-Planck equation for $N$ and $\theta$. We can do 
this by substitution of $\phi=\sqrt{N} e^{i \theta}$ into the 
action in the exponent of Eq.~(\ref{singlemodeprobdistr}). Since 
the jacobian of the transformation is equal to one, we simply have 
\begin{equation} 
\label{probdistrntheta} 
 P[N,\theta;t] = 
   \int_{\theta(t)=\theta}^{N(t)=N} d[N] d[\theta] 
     \exp 
     \left\{ 
       \frac{i}{\hbar} S [N,\theta] 
     \right\}, 
\end{equation} 
with an action equal to 
\begin{equation} 
\label{actionntheta} 
  S[N,\theta] = \int_{t_0}^t dt' 
   \left( 
     \frac{2 N(t')}{\hbar \Sigma^{\rm K}} 
       \left( 
         \hbar \dot \theta (t') + \mu_{\rm c} -\mu 
       \right)^2 
      + 
       \frac{\hbar}{2 \Sigma^{\rm K} N(t')} 
        \left( 
      \dot N (t') + \frac{2 R}{\hbar} N (t') 
    \right)^2 
   \right). 
\end{equation} 
Naively, we could derive the Fokker-Planck equation from the 
above path-integral expression  by going through the same steps 
as before. If we again apply normal ordering to the hamiltonian 
with respect to $N$ and $\theta$ and their conjugate momenta, the 
Fokker-Planck equation reads 
\begin{eqnarray} 
\label{wrongfpeqnntheta} 
  \frac{\partial P[N,\theta;t]}{\partial t} 
    &=&\left( \frac{i \Sigma^{\rm K}}{2} \frac{\partial^2}{\partial N^2} N + 
    \frac{2R}{\hbar} \frac{\partial}{\partial N} N \right) P[N,\theta;t] 
      \nonumber \\ 
    &+& 
     \left( 
      \frac{i \Sigma^{\rm K}}{8 N} \frac{\partial^2}{\partial \theta^2} 
      + \frac{1}{\hbar} \frac{\partial}{\partial \theta} (\mu_{\rm c} - \mu) 
    \right) P [N,\theta;t]. 
\end{eqnarray} 
This Fokker-Planck equation is however incorrect, since it is 
easily seen that the equilibrium distribution in 
Eq.~(\ref{equiln}) is not a solution of this Fokker-Planck 
equation. The fact that the Fokker-Planck equation in 
Eq.~(\ref{wrongfpeqnntheta}) turns out to be incorrect has to do 
with the fact that we have normal ordered the hamiltonian in 
terms of the variables $N$ and $\theta$. Although normal ordering 
of the hamiltonian $H[p_{\phi},\phi;p_{\phi^*},\phi^*]$ did give 
the correct results, this, however, does not imply that we also 
have to normal order the hamiltonian in terms of $N$ and 
$\theta$. Let us therefore proceed more carefully, and rewrite 
the Fokker-Planck equation in Eq.~(\ref{fpsinglemodephi}) for 
$P[\phi^*,\phi;t]$ in terms of $N$ and $\theta$. With the use of 
the chain rule for differentiation it is easy to show that, for a 
general function $f$ 
\begin{equation} 
  \frac{\partial f}{\partial \phi^*} = 
    \sqrt{N} e^{i \theta} \frac{\partial f}{\partial N} 
    + \frac{i}{2 \sqrt{N}} e^{i \theta} \frac{\partial f}{\partial \theta}, 
\end{equation} 
with the complex conjugate expression for $\partial f /\partial \phi$. 
Substitution of this result in the Fokker-Planck equation 
in Eq.~(\ref{fpsinglemodephi}) yields the Fokker-Planck equation for 
$P[N,\theta;t]$. It is given by 
\begin{eqnarray} 
\label{rightfpeqnntheta} 
  \frac{\partial P[N,\theta;t]}{\partial t} 
    &=&\left( \frac{i \Sigma^{\rm K}}{2} \frac{\partial}{\partial N} N 
    \frac{\partial}{\partial N}+ 
    \frac{2R}{\hbar} \frac{\partial}{\partial N} N \right) P[N,\theta;t] 
      \nonumber \\ 
    &+& 
    \left( 
      \frac{i \Sigma^{\rm K}}{8 N} \frac{\partial^2}{\partial \theta^2} 
      + \frac{1}{\hbar} \frac{\partial}{\partial \theta} (\mu_{\rm c} - \mu) 
    \right) P [N,\theta;t]. 
\end{eqnarray} 
Comparison of the Fokker-Planck equations in 
Eqs.~(\ref{wrongfpeqnntheta}) and (\ref{rightfpeqnntheta}) shows 
that in Eq.~(\ref{wrongfpeqnntheta}) we have misinterpreted the 
noise on $N(t)$ as an Ito process, whereas in the correct 
Fokker-Planck equation in Eq.~(\ref{rightfpeqnntheta}) we are 
clearly dealing with a Stratonovich process. Note that the same 
conclusion can also be reached by determining the equation of 
motion of $\langle N \rangle (t)$ from a variation of the action 
$S[N,\theta]$. 
 
From the action in Eq.~(\ref{actionntheta}) we can read of the Langevin equations 
for $N$ and $\theta$. The Langevin equation for $N(t)$ is given by 
\begin{eqnarray} 
\label{nlangevinsm} 
 \dot N (t) &=& -\frac{2R}{\hbar} N (t) + 2 \sqrt{N(t)} \eta (t') ; 
                \nonumber \\ 
 \langle \eta(t') \eta(t) \rangle &=& \frac{i \Sigma^{\rm K}}{4} \delta (t'-t), 
\end{eqnarray} 
and the Langevin equation for $\theta$ reads 
\begin{eqnarray} 
 \hbar \dot \theta (t) &=& \mu - \mu_{\rm c} + \frac{\nu (t)}{\sqrt{N(t)}}; 
                \nonumber \\ 
 \langle \nu (t') \nu (t) \rangle &=& \frac{i \hbar^2 \Sigma^{\rm K}}{4} \delta 
 (t'-t). 
\end{eqnarray} 
From the above discussion we thus conclude that we have to 
interpret the noise in the Langevin equation in 
Eq.~(\ref{nlangevinsm}) for the number of particles $N(t)$ as a 
Stratonovich process, to achieve the correct equilibrium 
distribution. This is the main conclusion of this appendix. It is 
straightforward to show that the above discussion generalizes to 
the case of a multi-mode description of the condensate in terms 
of the complex field $\phi \args$. When we make the 
transformation to density and phase variables by setting $\phi = 
\sqrt{\rho} e^{i \theta}$, we again have to be careful, and 
interpret the multiplicative noise that enters the equation of 
motion for the density $\rho \args$ as a Stratonovich process.

\end{document}